\documentclass[reprint,superscriptaddress]{revtex4-1}

\usepackage{graphicx}
\usepackage[english]{babel}
\usepackage{bbold}
\usepackage{braket}
\usepackage{graphicx}
\usepackage{amsthm}
\usepackage{float}
\usepackage{makeidx}
\usepackage{enumitem}
\usepackage{tikz}
\usepackage{url}
\usepackage[utf8]{inputenc}
\usepackage[bottom]{footmisc}
\usepackage{amsmath}
\usepackage{amssymb}
\usepackage[caption=false]{subfig}
\usepackage{scrextend}
\usepackage[colorlinks=true,citecolor=red,linkcolor=brown,urlcolor=magenta]{hyperref}

\newtheorem{theorem}{Theorem}

\DeclareMathAlphabet{\pazocal}{OMS}{zplm}{m}{n}
\newcommand{\E}{\pazocal{E}}

\newcommand{\F}{\pazocal{F}}
\newcommand{\R}{\pazocal{R}}

\newcommand{\G}{\pazocal{G}}
\newcommand{\C}{\pazocal{C}}
\newcommand{\p}{\pazocal{P}}

\newcommand{\Q}{\pazocal{Q}}

\newcommand{\U}{\pazocal{U}}
\newcommand{\M}{\pazocal{M}}

\newcommand{\Z}{\pazocal{Z}}
\newcommand{\X}{\pazocal{X}}
\newcommand{\Y}{\pazocal{Y}}
\newcommand{\I}{\pazocal{I}}

\newcommand{\dg}{\dagger}

\begin{document}

\title{Experimental accreditation of outputs of noisy quantum computers}

\author{Samuele Ferracin}
\email{samuele.ferracin@gmail.com}
\affiliation{Department of Applied Mathematics, University of Waterloo, Waterloo, Ontario N2L 3G1, Canada}
\affiliation{Department of Physics, University of Warwick, Coventry CV4 7AL, United Kingdom}

\author{Seth T. Merkel}
\email{seth.merkel@ibm.com}
\affiliation{IBM Quantum, T.J. Watson Research Center, Yorktown Heights, NY 10598, USA}

\author{David McKay}
\email{dcmckay@us.ibm.com}
\affiliation{IBM Quantum, T.J. Watson Research Center, Yorktown Heights, NY 10598, USA}

\author{Animesh Datta}
\email{animesh.datta@warwick.ac.uk}
\affiliation{Department of Physics, University of Warwick, Coventry CV4 7AL, United Kingdom}
\begin{abstract}
We provide and experimentally demonstrate an accreditation protocol that upper-bounds the variation distance between noisy and noiseless probability distributions of the outputs of arbitrary quantum computations. We accredit the outputs of twenty-four quantum circuits executed on programmable superconducting hardware, ranging from depth nine circuits on ten qubits to depth twenty-one circuits on four qubits. Our protocol requires implementing the ``target'' quantum circuit along with a number of random Clifford circuits and subsequently post-processing the outputs of these Clifford circuits. Importantly, the number of Clifford circuits is chosen to obtain the bound with the desired confidence and accuracy, and is independent of the size and nature of the target circuit. We thus demonstrate a practical and scalable method of ascertaining the correctness of the outputs of arbitrary-sized noisy quantum computers—the ultimate arbiter of the utility of the computer itself.
\end{abstract}
\date{\today}
\maketitle
\twocolumngrid

\textit{1. Introduction|}
The utility of noisy quantum computers in simulation and optimisation will be determined by our ability to ascertain if the solutions provided are correct or close to correct. 
This is a challenging task for problems that are outside the complexity class NP. The current methods are based on evaluating single-valued metrics such as the Cross Entropy \cite{B&al16,GoogleSupremacy19} or the Quantum Volume \cite{CBSNG19}, which can be linked to the performance of the quantum hardware being used. These methods require simulating the relevant quantum circuits on classical  computers.   Though  practical  at  present,  they are not scalable and consequently useless for problems that cannot already be simulated classically.  On the contrary, the proposals based on quantum cryptography and interactive proof systems are scalable in principle,  but have an overhead in width (qubits) and depth (gates) that makes them impractical for the foreseeable future~\cite{C05,FK12,RUV12,B15,HM15,MF16,FKD17,UM18}. This calls for new methods that are both practical in the short term and scalable in the long term.

In this work we present and experimentally demonstrate an Accreditation Protocol (AP) that achieves this goal.  This AP provides an upper bound on the variation distance (VD) between the probability distribution of the experimental outputs of a noisy quantum circuit $\{p_{\textup{exp}}(\overline{s})\}$ and its ideal, noiseless counterpart $\{p_{\textup{ideal}}(\overline{s})\}$, where $\overline{s}$ denotes the bit strings that may be obtained as output. In our AP, the ``target'' quantum circuit  the correctness of whose outputs we wish to ascertain is executed along with a number $v$ of random Clifford circuits (the ``traps''). The trap circuits have the same width and depth as the target circuit and are designed such that in the \textit{absence} of noise they always return a fixed known output.  This enables us,  in the \textit{presence} of noise,  to measure the probability $p_{\textup{inc}}$ that a trap's output is incorrect. Our AP guarantees that the VD between $\{p_{\textup{exp}}(\overline{s})\}$ and $\{p_{\textup{ideal}}(\overline{s})\}$ is bounded as (see section \hyperlink{sec:proof_bound}1 of the Appendix for a proof)
\begin{equation}
\label{eq:boundVD}
\textrm{VD}:=\frac{1}{2}\sum_{\overline{s}}\big|p_{\textup{ideal}}(\overline{s})-p_{\textup{exp}}(\overline{s})\big|\leq 2p_{\textup{inc}}.
\end{equation}
The value $2p_{\textup{inc}}$ is estimated experimentally with accuracy $\theta\in(0,1)$ and confidence $\alpha\in(0,1)$ chosen by the user.  The number $v$ of trap circuits is determined by the desired $\theta$ and $\alpha$, but is independent of the size and nature of the target circuit~(Eq. \ref{eq:hoeff}).

\begin{figure}
     \centering
     \subfloat[][Accreditation of GHZ and QFT circuits.]{\includegraphics[clip,width=0.425\columnwidth]{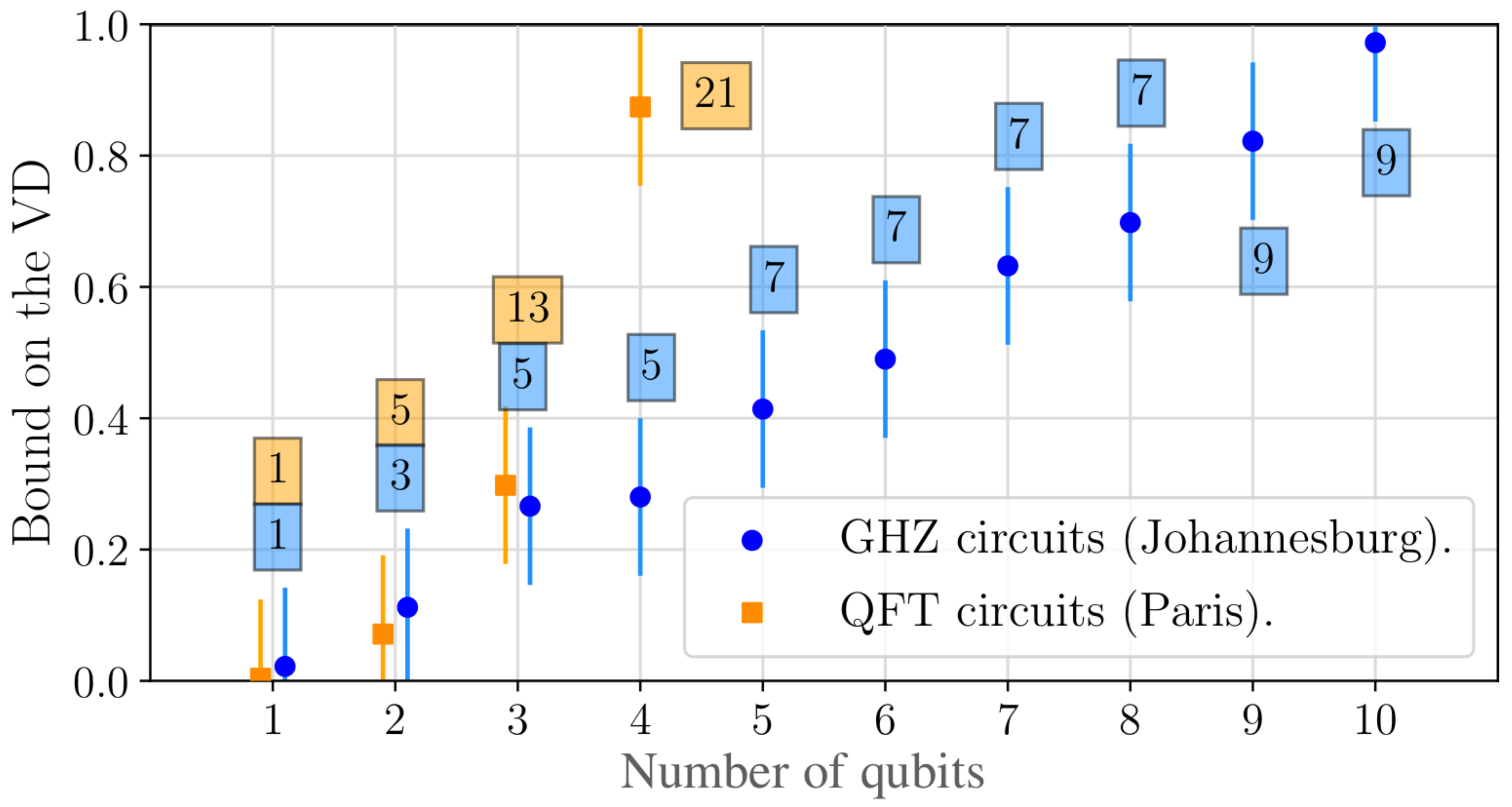}\label{fig:f1}}
     \qquad
     \qquad
     \subfloat[][Accreditation of six-qubit random circuits.]{\includegraphics[clip,width=0.41\columnwidth]{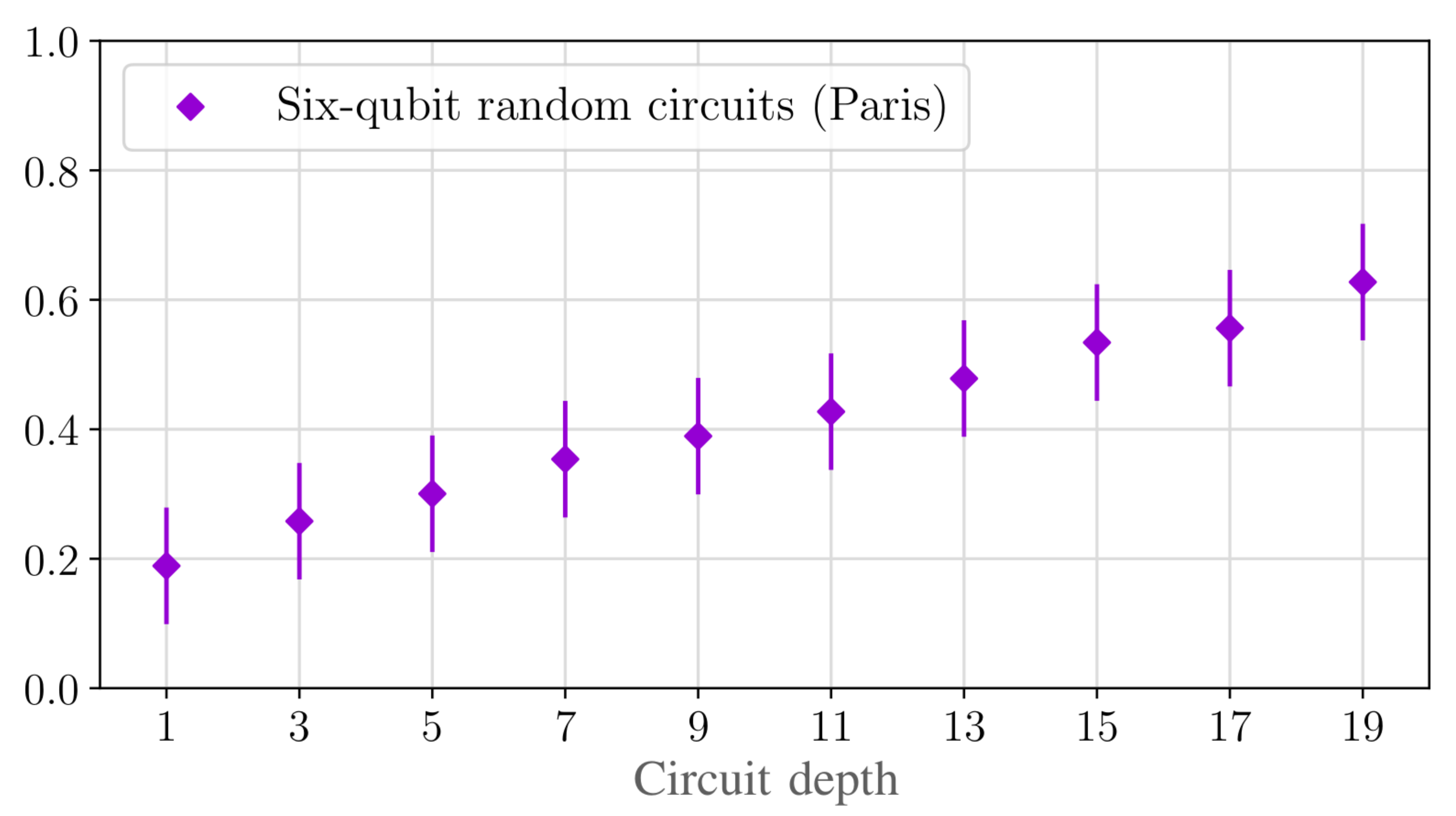}\label{fig:f2}}
     \caption{\small Experimental bounds on the VD (r.h.s. of Eq.  \ref{eq:boundVD}) provided by our AP. In {(a)} the numbers inside the rectangles indicate the depths of the various target circuits. The bounds in {(a)} are calculated by implementing $v=450$ trap circuits, those in {(b)} by implementing $v=900$ trap circuits. The bars correspond to confidence levels above 95$\%$ on our estimates of $2p_{\textrm{inc}}$|specifically, in (a) we set $\theta=13\%$ and $\alpha=95\%$, in (b) $\theta=9\%$ and $\alpha=95\%$,  see Eq. \ref{eq:hoeff}.}
     \label{fig:bounds_best_intro}
\end{figure}

We implement our AP on $\textsf{ibmq}\_\textsf{johannesburg}$ and $\textsf{ibmq}\_\textsf{paris}$, two two-dimensional arrays of superconducting transmon qubits. Fig.~\ref{fig:bounds_best_intro} shows the bounds provided by our AP for twenty-four different circuits.
Of these, fourteen are structured circuits|ten circuits to generate Greenberger-Horne-Zeilinger (GHZ) states~\cite{GHZ89} and four to perform the quantum Fourier transform~(QFT) \cite{NC00}, both important primitives in quantum computation|and ten are six-qubit random circuits of varying depth. The widest of these circuits has ten qubits and depth nine, the deepest has four qubits and depth twenty-one. The widths of our circuits compare favourably to that reached in the experimental demonstrations of some of the main protocols for noise characterization|three qubits in Process Tomography~\cite{WHEB04}, five in Randomized Benchmarking \cite{PCDRNBKY19}, seven in Direct Fidelity Estimation \cite{Luandothers15} and ten in Cycle Benchmarking~\cite{Erhard&al19}. 

Our AP is designed to ascertain the correctness of a noisy quantum computation rather than the performance of individual gates or families thereof. Thus, it can detect noise (such as location-dependant noise acting on the whole register of qubits) that may arise when gates are put together to form a circuit and may be missed by the protocols for characterizing individual gates~\cite{WHEB04,MGSPC13,
BK&al13,
BK&al17,Merkelandothers13,BK&al13, Luandothers15,
FL11,
HFW19,FW19,Erhard&al19}. An alternate accreditation protocol can detect even more complex noise such as temporally-correlated qubit-environment couplings, albeit at the cost of looser bounds

\newpage
\onecolumngrid

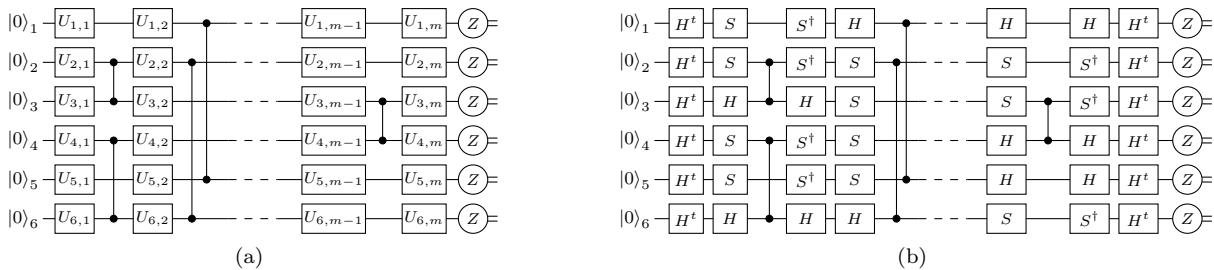
\begin{figure}
     \centering
     \subfloat[][]{

\begin{tikzpicture}[scale=0.65, every node/.style={scale=0.85}]

\foreach \x in {1,...,6}
\node at (-0.4,4.8-\x*0.8) {\footnotesize $\ket{0}_{\x}$};

\foreach \x in {0,...,5}
\draw (-0.0,4.0-\x*0.8) -- (3.8,4.0-\x*0.8);

\foreach \x in {0,...,5}
\draw [dashed] (3.8,4.0-\x*0.8) -- (4.8,4.0-\x*0.8);

\foreach \x in {0,...,5}
\draw (4.8,4.0-\x*0.8) -- (9.0,4.0-\x*0.8);

\foreach \x in {0,...,5}
\draw [fill=white] (0.25,4.0-\x*0.8-0.3) rectangle (1.05,4.0-\x*0.8+0.3);

\foreach \x in {1,...,6}
\node at (0.65,4.8-\x*0.8) {\scriptsize $U_{\x,1}$};

\draw [fill=black] (1.45,0.0) circle [radius=0.07cm];
\draw [fill=black] (1.45,1.6) circle [radius=0.07cm];
\draw [fill=black] (1.45,2.4) circle [radius=0.07cm];
\draw [fill=black] (1.45,3.2) circle [radius=0.07cm];

\draw (1.45,0.0) -- (1.45,1.6);
\draw (1.45,3.2) -- (1.45,2.4);

\foreach \x in {0,...,5}
\draw [fill=white] (0.25+1*1.6,4.0-\x*0.8-0.3) rectangle (1.05+1*1.6,4.0-\x*0.8+0.3);

\node at (0.65+1*1.6,4.8-1*0.8) {\scriptsize $U_{1,2}$};
\node at (0.65+1*1.6,4.8-2*0.8) {\scriptsize $U_{2,2}$};
\node at (0.65+1*1.6,4.8-3*0.8) {\scriptsize $U_{3,2}$};
\node at (0.65+1*1.6,4.8-4*0.8) {\scriptsize $U_{4,2}$};
\node at (0.65+1*1.6,4.8-5*0.8) {\scriptsize $U_{5,2}$};
\node at (0.65+1*1.6,4.8-6*0.8) {\scriptsize $U_{6,2}$};

\draw [fill=black] (1.75+1*1.6,4.8-1*0.8) circle [radius=0.07cm];
\draw [fill=black] (1.45+1*1.6,4.8-2*0.8) circle [radius=0.07cm];
\draw [fill=black] (1.75+1*1.6,4.8-5*0.8) circle [radius=0.07cm];
\draw [fill=black] (1.45+1*1.6,4.8-6*0.8) circle [radius=0.07cm];

\draw (1.45+1*1.6,0.0) -- (1.45+1*1.6,3.2);
\draw (1.75+1*1.6,4.0) -- (1.75+1*1.6,0.8);



\foreach \x in {0,...,5}
\draw [fill=white] (5.30,4.0-\x*0.8-0.3) rectangle (6.6,4.0-\x*0.8+0.3);

\node at (5.4+0.55,4.8-1*0.8) {\scriptsize $U_{1,m-1}$};
\node at (5.4+0.55,4.8-2*0.8) {\scriptsize $U_{2,m-1}$};
\node at (5.4+0.55,4.8-3*0.8) {\scriptsize $U_{3,m-1}$};
\node at (5.4+0.55,4.8-4*0.8) {\scriptsize $U_{4,m-1}$};
\node at (5.4+0.55,4.8-5*0.8) {\scriptsize $U_{5,m-1}$};
\node at (5.4+0.55,4.8-6*0.8) {\scriptsize $U_{6,m-1}$};

\draw [fill=black] (6.95,4.8-3*0.8) circle [radius=0.07cm];
\draw [fill=black] (6.95,4.8-4*0.8) circle [radius=0.07cm];

\draw (6.95,1.6) -- (6.95,2.4);


\foreach \x in {0,...,5}
\draw [fill=white] (7.35,4.0-\x*0.8-0.3) rectangle (8.25,4.0-\x*0.8+0.3);

\foreach \x in {1,...,6}
\node at (7.4+0.4,4.8-\x*0.8) {\scriptsize $U_{\x,m}$};

\foreach \x in {1,...,6}
\draw (9.1,4.8-\x*0.8-0.05) -- (9.3,4.8-\x*0.8-0.05);
\foreach \x in {1,...,6}
\draw (9.1,4.8-\x*0.8+0.05) -- (9.3,4.8-\x*0.8+0.05);

\foreach \x in {0,...,5}
\draw [fill=white] (8.8,\x*0.8) circle [radius=0.3cm];
\foreach \x in {1,...,6}
\node at (8.8,4.8-\x*0.8) {\scriptsize $Z$};

\end{tikzpicture}
\label{fig:target}
     }
     \qquad
     \qquad
     \subfloat[][]{

\begin{tikzpicture}[scale=0.65, every node/.style={scale=0.85}]

\foreach \x in {1,...,6}
\node at (-0.3,4.8-\x*0.8) {\footnotesize $\ket{0}_{\x}$};

\foreach \x in {0,...,5}
\draw (0.1,4.0-\x*0.8) -- (5.5,4.0-\x*0.8);

\foreach \x in {0,...,5}
\draw [dashed] (5.5,4.0-\x*0.8) -- (6.5,4.0-\x*0.8);

\foreach \x in {0,...,5}
\draw (11,4.0-\x*0.8) -- (6.5,4.0-\x*0.8);

\foreach \x in {0,...,5}
\draw [fill=white] (0.4,4.0-\x*0.8-0.3) rectangle (1.1,4.0-\x*0.8+0.3);

\foreach \x in {1,...,6}
\node at (0.75,4.8-\x*0.8) {\scriptsize $H^t$};

\foreach \x in {0,...,5}
\draw [fill=white] (1.3,4.0-\x*0.8-0.3) rectangle (2.0,4.0-\x*0.8+0.3);

\node at (1.65,4.8-1*0.8) {\scriptsize $S$};
\node at (1.65,4.8-2*0.8) {\scriptsize $S$};
\node at (1.65,4.8-3*0.8) {\scriptsize $H$};
\node at (1.65,4.8-4*0.8) {\scriptsize $S$};
\node at (1.65,4.8-5*0.8) {\scriptsize $S$};
\node at (1.65,4.8-6*0.8) {\scriptsize $H$};

\draw [fill=black] (2.45,4.8-2*0.8) circle [radius=0.07cm];
\draw [fill=black] (2.45,4.8-3*0.8) circle [radius=0.07cm];
\draw [fill=black] (2.45,4.8-4*0.8) circle [radius=0.07cm];
\draw [fill=black] (2.45,4.8-6*0.8) circle [radius=0.07cm];

\draw (2.45,4.8-2*0.8) -- (2.45,4.8-3*0.8);
\draw (2.45,4.8-4*0.8) -- (2.45,4.8-6*0.8);

\foreach \x in {0,...,5}
\draw [fill=white] (2.8,4.0-\x*0.8-0.3) rectangle (3.6,4.0-\x*0.8+0.3);

\node at (3.2,4.8-1*0.8) {\scriptsize $S^\dagger$};
\node at (3.2,4.8-2*0.8) {\scriptsize $S^\dg$};
\node at (3.2,4.8-3*0.8) {\scriptsize $H$};
\node at (3.2,4.8-4*0.8) {\scriptsize $S^\dagger$};
\node at (3.2,4.8-5*0.8) {\scriptsize $S^\dagger$};
\node at (3.2,4.8-6*0.8) {\scriptsize $H$};

\foreach \x in {0,...,5}
\draw [fill=white] (3.8,4.0-\x*0.8-0.3) rectangle (4.6,4.0-\x*0.8+0.3);

\node at (4.2,4.8-1*0.8) {\scriptsize $H$};
\node at (4.2,4.8-2*0.8) {\scriptsize $S$};
\node at (4.2,4.8-3*0.8) {\scriptsize $S$};
\node at (4.2,4.8-4*0.8) {\scriptsize $S$};
\node at (4.2,4.8-5*0.8) {\scriptsize $S$};
\node at (4.2,4.8-6*0.8) {\scriptsize $H$};

\draw [fill=black] (5.25,4.8-1*0.8) circle [radius=0.07cm];
\draw [fill=black] (4.9+.15,4.8-2*0.8) circle [radius=0.07cm];
\draw [fill=black] (5.25,4.8-5*0.8) circle [radius=0.07cm];
\draw [fill=black] (4.9+.15,4.8-6*0.8) circle [radius=0.07cm];

\draw (4.9+.15,0.0) -- (4.9+.15,3.2);
\draw (5.25,4.0) -- (5.25,0.8);



\foreach \x in {0,...,5}
\draw [fill=white] (6.9,4.0-\x*0.8-0.3) rectangle (7.7,4.0-\x*0.8+0.3);

\node at (7.3,4.8-1*0.8) {\scriptsize $H$};
\node at (7.3,4.8-2*0.8) {\scriptsize $S$};
\node at (7.3,4.8-3*0.8) {\scriptsize $S$};
\node at (7.3,4.8-4*0.8) {\scriptsize $H$};
\node at (7.3,4.8-5*0.8) {\scriptsize $H$};
\node at (7.3,4.8-6*0.8) {\scriptsize $S$};

\draw [fill=black] (8+.15,4.8-3*0.8) circle [radius=0.07cm];
\draw [fill=black] (8+.15,4.8-4*0.8) circle [radius=0.07cm];

\draw (8+.15,1.6) -- (8+.15,2.4);


\foreach \x in {0,...,5}
\draw [fill=white] (8.6,4.0-\x*0.8-0.3) rectangle (9.4,4.0-\x*0.8+0.3);

\node at (9,4.8-1*0.8) {\scriptsize $H$};
\node at (9,4.8-2*0.8) {\scriptsize $S^\dagger$};
\node at (9,4.8-3*0.8) {\scriptsize $S^\dagger$};
\node at (9,4.8-4*0.8) {\scriptsize $H$};
\node at (9,4.8-5*0.8) {\scriptsize $H$};
\node at (9,4.8-6*0.8) {\scriptsize $S^\dg$};

\foreach \x in {0,...,5}
\draw [fill=white] (9.6,4.0-\x*0.8-0.3) rectangle (10.4,4.0-\x*0.8+0.3);

\foreach \x in {0,...,5}
\node at (10,4.0-\x*0.8) {\scriptsize $H^t$};

\foreach \x in {1,...,6}
\draw (9.1+2.2,4.8-\x*0.8-0.05) -- (9.3+2.2,4.8-\x*0.8-0.05);
\foreach \x in {1,...,6}
\draw (9.1+2.2,4.8-\x*0.8+0.05) -- (9.3+2.2,4.8-\x*0.8+0.05);

\foreach \x in {0,...,5}
\draw [fill=white] (11,\x*0.8) circle [radius=0.3cm];
\foreach \x in {1,...,6}
\node at (9.8+1.2,4.8-\x*0.8) {\scriptsize $Z$};

\end{tikzpicture}
\label{fig:trap}
     }
     \caption{\small (a) Example of target circuit.  The target circuit must be compiled into $m$ cycles of one-qubit gates,  each one (apart from the last one) followed by a cycle of $cZ$ gates (giving circuit depth $d=2m-1$).  Input qubits are in the state $\ket{0}$ and measurements are in the Pauli-$Z$ basis.  (b) Example of trap circuit for the target circuit in (a). The trap circuit is obtained by replacing the one-qubit gates in the target circuit with one-qubit Clifford gates.  Neighboring cycles of one-qubit gates can be recompiled into a single cycle. Thus, the trap circuit has the same circuit depth as the target. }
     \label{fig:circuits_examples}
\end{figure}
\twocolumngrid

\noindent on the VD~\cite{FKD18}|more details in section \hyperlink{sec:comparison}3 in the Appendix.  Due to its practicality, scalability and ability to capture a broad class of noise processes,we expect that in the future our AP will supplant the protocols based on classical simulations of quantum circuits.

We begin in section~\hyperlink{sec:notation}2 by introducing the notation and the noise model, in section~\hyperlink{sec:AP}3 we present our AP, in sections~\hyperlink{sec:exp}4 and \hyperlink{sec:diagnostic}5 we discuss the experimental results.

\hypertarget{sec:notation}{\textit{2. Notation and noise model|}}We indicate unitaries with capital letters and Completely Positive Trace Preserving (CPTP) maps with calligraphic letters. We use $I=\textrm{diag(1,1)}$ to denote the identity, $X$, $Y$ and $Z$ for the one-qubit Pauli matrices, $H=(X+Z)/\sqrt{2}$ for the Hadamard gate, $S=|0\rangle\langle0|+i|1\rangle\langle1|$ for the phase gate, $cZ=|0\rangle\langle0|\otimes I+|1\rangle\langle1|\otimes Z$ for the controlled-$Z$ gate and $cX=|0\rangle\langle0|\otimes I+|1\rangle\langle1|\otimes X$ for the controlled-$X$ gate. We indicate with ``cycle'' a set of gates acting on the entire system within a fixed period of time.

We model noise in state preparation, measurements and cycles as CPTP maps acting on all the qubits. Specifically, we assume that a noisy implementation of a cycle $\G$ on a state $\rho$ at circuit depth $j$ returns $\E_{\G,j}\G(\rho)$, where $\E_{\G,j}$ is a CPTP map that potentially acts on the whole system and depends on both $\G$ and on the depth~$j$. This is a Markovian noise model that encompasses a broad class of noise processes afflicting current platforms, e.g.,  gate-dependent noise and cross-talk. It is more general than the noise models typically considered in the protocols for gate characterization, where the noise is represented by a static map $\E_\G$ independent of~$j$~\cite{WHEB04,MGSPC13,
BK&al13,
BK&al17,Merkelandothers13,BK&al13, Luandothers15,
FL11,HFW19,FW19,Erhard&al19,LH20}.

We assume that the cycles of one-qubit gates suffer gate-independent noise, i.e. $\E_{\U,j}=\E_j$ for all the cycles of one-qubit gates $\U$. In our analysis this assumption is required for two reasons: Firstly, to transform arbitrary noise processes into Pauli errors via a quantum one-time pad (QOTP, see section \hyperlink{sec:AP}3), and secondly, to ensure that the distributions of errors afflicting target and traps are identical. This is a common assumption in the literature on noise characterization protocols \cite{CN98,WHEB04,MGSPC13,BK&al13,Merkelandothers13,DanG15,BK&al17,FL11,DLP11,MDRL12,Luandothers15,
K&al07,MGE11,Erhard&al19,HFW19,
LH20} and is motivated by the empirical observation that the one-qubit gates are the most accurate components in all the leading platforms~\cite{WrightEtal19,GoogleSupremacy19}. Nevertheless, we relax it by showing that the
bound provided by our AP is robust to

\newpage
\onecolumngrid

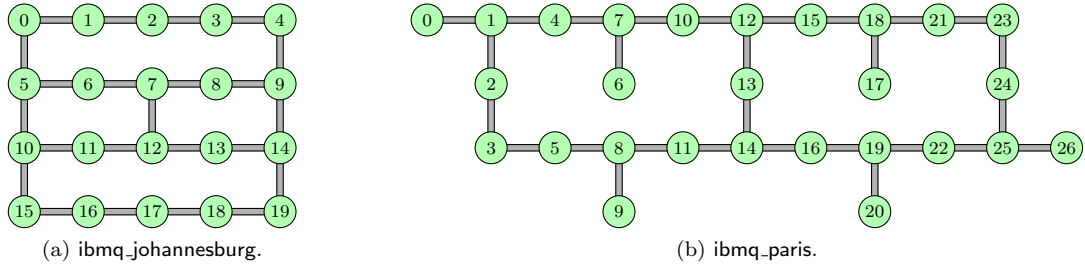
\begin{figure}[H]
     \centering
     \subfloat[][$\textsf{ibmq}\_\textsf{johannesburg}$.]{
     
     \begin{tikzpicture}[scale=0.85, every node/.style={scale=0.85}]
           
      \foreach \x in {0,4} 
      \draw [gray!60, line width=1mm](\x, 0) -- (\x, 3);
      \foreach \x in {2} 
      \draw [gray!60, line width=1mm](\x, 1) -- (\x, 2);
      \foreach \y in {0,1,2,3} 
      \draw [gray!60, line width=1mm](0,\y) -- (4, \y);
      
       \foreach \x in {0,4} 
      \draw (\x-0.05, 0) -- (\x-0.05, 3);
      \foreach \x in {0,4} 
      \draw (\x+0.05, 0) -- (\x+0.05, 3);
      \foreach \x in {2} 
      \draw (\x-0.05, 1) -- (\x-0.05, 2);
      \foreach \x in {2} 
      \draw (\x+0.05, 1) -- (\x+0.05, 2);
      \foreach \y in {0,1,2,3} 
      \draw (0,\y+0.05) -- (4, \y+0.05);
      \foreach \y in {0,1,2,3} 
      \draw (0,\y-0.05) -- (4, \y-0.05);
      
  \foreach \x in {0,1,2,3, 4} 
    \foreach \y in {0,1,2,3} 
      { 
        \draw [fill=green!30] (\x,\y) circle (0.25cm);
      } 
      
      \node at (0,3) {\footnotesize 0};
      \node at (1,3) {\footnotesize 1};
      \node at (2,3) {\footnotesize 2};
      \node at (3,3) {\footnotesize 3};
      \node at (4,3) {\footnotesize 4};
      \node at (0,2) {\footnotesize 5};
      \node at (1,2) {\footnotesize 6};
      \node at (2,2) {\footnotesize 7};
      \node at (3,2) {\footnotesize 8};
      \node at (4,2) {\footnotesize 9};
      \node at (0,1) {\footnotesize 10};
      \node at (1,1) {\footnotesize 11};
      \node at (2,1) {\footnotesize 12};
      \node at (3,1) {\footnotesize 13};
      \node at (4,1) {\footnotesize 14};
      \node at (0,0) {\footnotesize 15};
      \node at (1,0) {\footnotesize 16};
      \node at (2,0) {\footnotesize 17};
      \node at (3,0) {\footnotesize 18};
      \node at (4,0) {\footnotesize 19};
      
\end{tikzpicture}
     
     }
     \qquad
     \qquad
     \subfloat[][$\textsf{ibmq}\_\textsf{paris}$.]{

     \begin{tikzpicture}[scale=0.85, every node/.style={scale=0.85}]
      
      \draw [gray!60, line width=1mm](1, 1) -- (1, 3); \draw (1-0.05, 1) -- (1-.05, 3); \draw (1.05, 1) -- (1.05, 3);
      \draw [fill=green!30] (1,2) circle (0.25cm);
      \node at (1,2) {\footnotesize 2};
      \draw[gray!60, line width=1mm] (5, 1) -- (5, 3);
      \draw (5.05, 1) -- (5.05, 3);
      \draw (5-.05, 1) -- (5-.05, 3);
      \draw [fill=green!30] (5,2) circle (0.25cm);
      \node at (5,2) {\footnotesize 13};
      \draw [gray!60, line width=1mm](9, 1) -- (9, 3);
      \draw (9.05, 1) -- (9.05, 3);
      \draw (9-.05, 1) -- (9-.05, 3); 
      \draw [fill=green!30] (9,2) circle (0.25cm);
      \node at (9,2) {\footnotesize 24};
      \draw [gray!60, line width=1mm](3, 3) -- (3, 2);
      \draw (3.05, 3) -- (3.05, 2);
      \draw(3-.05, 3) -- (3-.05, 2);
      \draw [fill=green!30] (3,2) circle (0.25cm);
      \node at (3,2) {\footnotesize 6};
      \draw [gray!60, line width=1mm](7, 3) -- (7, 2);
      \draw(7.05, 3) -- (7.05, 2);
      \draw(7-.05, 3) -- (7-.05, 2);
      \draw [fill=green!30] (7,2) circle (0.25cm);
      \node at (7,2) {\footnotesize 17};
      \draw [gray!60, line width=1mm](3, 1) -- (3, 0);
      \draw(3.05, 1) -- (3.05, 0);
      \draw(3-.05, 1) -- (3-.05, 0);
      \draw [fill=green!30] (3,0) circle (0.25cm);
      \node at (3,0) {\footnotesize 9};
      \draw [gray!60, line width=1mm](7, 1) -- (7, 0);
      \draw(7.05, 1) -- (7.05, 0);
      \draw(7-.05, 1) -- (7-.05, 0);
      \draw [fill=green!30] (7,0) circle (0.25cm);
      \node at (7,0) {\footnotesize 20};
           
      \draw [gray!60, line width=1mm] (0, 3) -- (9, 3);
      \draw(0, 3.05) -- (9, 3.05);
      \draw(0, 3-.05) -- (9, 3-.05);
  \foreach \x in {0,1,2,3, 4,5,6,7,8,9} 
        \draw [fill=green!30] (\x,3) circle (0.25cm);
      
      \node at (0,3) {\footnotesize 0};
      \node at (1,3) {\footnotesize 1};
      \node at (2,3) {\footnotesize 4};
      \node at (3,3) {\footnotesize 7};
      \node at (4,3) {\footnotesize 10};
      \node at (5,3) {\footnotesize 12};
      \node at (6,3) {\footnotesize 15};
      \node at (7,3) {\footnotesize 18};
      \node at (8,3) {\footnotesize 21};
      \node at (9,3) {\footnotesize 23};
      
      \draw [gray!60, line width=1mm](1, 1) -- (10, 1);
      \draw(1, 1.05) -- (10, 1.05);
      \draw(1, 1-.05) -- (10, 1-.05);
  \foreach \x in {10,1,2,3, 4,5,6,7,8,9} 
        \draw [fill=green!30] (\x,1) circle (0.25cm);
      
      \node at (1,1) {\footnotesize 3};
      \node at (2,1) {\footnotesize 5};
      \node at (3,1) {\footnotesize 8};
      \node at (4,1) {\footnotesize 11};
      \node at (5,1) {\footnotesize 14};
      \node at (6,1) {\footnotesize 16};
      \node at (7,1) {\footnotesize 19};
      \node at (8,1) {\footnotesize 22};
      \node at (9,1) {\footnotesize 25};
      \node at (10,1) {\footnotesize 26};
      
\end{tikzpicture}

     }
     \caption{\small A graphical representation of the connectivity in $\textsf{ibmq}\_\textsf{johannesburg}$ and $\textsf{ibmq}\_\textsf{paris}$. The circles represent qubits,  the edges represent the available entangling gates.}
     \label{fig:ibmq_architecture}
\end{figure}

\twocolumngrid

\begin{figure}[H]
     \centering
     \subfloat[][Our largest GHZ circuit.]{
     \begin{tikzpicture}[scale=0.65, every node/.style={scale=0.85}]

\draw [white] (0.,9.8*0.8) -- (1.35+0.3+11*0.6,9.8*0.8);

\node at (-0.4,9*0.8) {\footnotesize $\ket{0}_{5}$};
\node at (-0.4,8*0.8) {\footnotesize $\ket{0}_{10}$};
\node at (-0.4,7*0.8) {\footnotesize $\ket{0}_{0}$};
\node at (-0.4,6*0.8) {\footnotesize $\ket{0}_{15}$};
\node at (-0.4,5*0.8) {\footnotesize $\ket{0}_{1}$};
\node at (-0.4,4*0.8) {\footnotesize $\ket{0}_{6}$};
\node at (-0.4,3*0.8) {\footnotesize $\ket{0}_{11}$};
\node at (-0.4,2*0.8) {\footnotesize $\ket{0}_{16}$};
\node at (-0.4,1*0.8) {\footnotesize $\ket{0}_{2}$};
\node at (-0.4,0*0.8) {\footnotesize $\ket{0}_{7}$};
\node at (-0.4,0*0.8) {\footnotesize $\ket{0}_{7}$};

\foreach \x in {0,...,9}
\draw (0.,\x*0.8) -- (1.35+0.3+11*0.6,\x*0.8);

\foreach \x in {0,...,9}
\draw [fill=white] (0.3,\x*0.8-0.3) rectangle (1.0,\x*0.8+0.3);
\foreach \x in {0,...,9}
\node at (0.65,\x*0.8) {\scriptsize $H$};

\draw [fill=black] (1.35,9*0.8) circle [radius=0.07cm];
\draw [fill=black] (1.35,8*0.8) circle [radius=0.07cm];
\draw (1.35,9*0.8) -- (1.35,8*0.8-0.);

\draw [fill=white] (1.05+1*0.6,8*0.8-0.3) rectangle (1.05+0.7+1*0.6,8*0.8+0.3);
\node at (1.05+1*0.95,8*0.8) {\scriptsize $H$};

\draw [fill=black] (1.35+2*0.6,9*0.8) circle [radius=0.07cm];
\draw [fill=black] (1.35+2*0.6,7*0.8) circle [radius=0.07cm];
\draw (1.35+2*0.6,9*0.8) -- (1.35+2*0.6,7*0.8);

\draw [fill=black] (1.35+2.5*0.6,8*0.8) circle [radius=0.07cm];
\draw [fill=black] (1.35+2.5*0.6,6*0.8) circle [radius=0.07cm];
\draw (1.35+2.5*0.6,8*0.8) -- (1.35+2.5*0.6,6*0.8-0.);

\draw [fill=white] (1.05+3.5*0.6,7*0.8-0.3) rectangle (1.05+0.7+3.5*0.6,7*0.8+0.3);
\node at (1.05+0.35+3.5*0.6,7*0.8) {\scriptsize $H$};
\draw [fill=white] (1.05+3.5*0.6,6*0.8-0.3) rectangle (1.05+0.7+3.5*0.6,6*0.8+0.3);
\node at (1.05+0.35+3.5*0.6,6*0.8) {\scriptsize $H$};

\draw [fill=black] (1.35+4.7*0.6,7*0.8) circle [radius=0.07cm];
\draw [fill=black] (1.35+4.7*0.6,5*0.8) circle [radius=0.07cm];
\draw (1.35+4.7*0.6,7*0.8) -- (1.35+4.7*0.6,5*0.8-0.);

\draw [fill=black] (1.35+5.2*0.6,9*0.8) circle [radius=0.07cm];
\draw [fill=black](1.35+5.2*0.6,4*0.8) circle [radius=0.07cm];
\draw (1.35+5.2*0.6,9*0.8) -- (1.35+5.2*0.6,4*0.8-0.);

\draw [fill=black] (1.35+5.7*0.6,8*0.8) circle [radius=0.07cm];
\draw [fill=black] (1.35+5.7*0.6,3*0.8) circle [radius=0.07cm];
\draw (1.35+5.7*0.6,8*0.8) -- (1.35+5.7*0.6,3*0.8-0.);

\draw [fill=black] (1.35+6.2*0.6,6*0.8) circle [radius=0.07cm];
\draw [fill=black] (1.35+6.2*0.6,2*0.8) circle [radius=0.07cm];
\draw (1.35+6.2*0.6,6*0.8) -- (1.35+6.2*0.6,2*0.8-0.);

\draw [fill=white] (1.05+7.2*0.6,5*0.8-0.3) rectangle (1.05+0.7+7.2*0.6,5*0.8+0.3);
\node at (1.05+0.35+7.2*0.6,5*0.8) {\scriptsize $H$};
\draw [fill=white] (1.05+7.2*0.6,4*0.8-0.3) rectangle (1.05+0.7+7.2*0.6,4*0.8+0.3);
\node at (1.05+0.35+7.2*0.6,4*0.8) {\scriptsize $H$};
\draw [fill=white] (1.05+7.2*0.6,3*0.8-0.3) rectangle (1.05+0.7+7.2*0.6,3*0.8+0.3);
\node at (1.05+0.35+7.2*0.6,3*0.8) {\scriptsize $H$};
\draw [fill=white] (1.05+7.2*0.6,2*0.8-0.3) rectangle (1.05+0.7+7.2*0.6,2*0.8+0.3);
\node at (1.05+0.35+7.2*0.6,2*0.8) {\scriptsize $H$};

\draw [fill=black] (2.05+7.2*0.6,5*0.8) circle [radius=0.07cm];
\draw [fill=black] (2.05+7.2*0.6,1*0.8) circle [radius=0.07cm];
\draw (2.05+7.2*0.6,5*0.8) -- (2.05+7.2*0.6,1*0.8-0.);

\draw [fill=black] (2.05+7.7*0.6,4*0.8) circle [radius=0.07cm];
\draw [fill=black] (2.05+7.7*0.6,0*0.8) circle [radius=0.07cm];
\draw (2.05+7.7*0.6,4*0.8) -- (2.05+7.7*0.6,0*0.8-0.);

\draw [fill=white] (2.05+8.2*0.6,1*0.8-0.3) rectangle (2.05+8.2*0.6+0.7,1*0.8+0.3);
\node at (2.4+8.2*0.6,1*0.8) {\scriptsize $H$};
\draw [fill=white] (2.05+8.2*0.6,0*0.8-0.3) rectangle (2.05+8.2*0.6+0.7,0*0.8+0.3);
\node at (2.4+8.2*0.6,0*0.8) {\scriptsize $H$};

\foreach \x in {0,...,9}
\draw (1.35+11*0.6+0.3,\x*0.8-0.05) -- (1.35+12*0.6+0.3,\x*0.8-0.05);
\foreach \x in {0,...,9}
\draw (1.35+11*0.6+0.3,\x*0.8+0.05) -- (1.35+12*0.6+0.3,\x*0.8+0.05);

\foreach \x in {0,...,9}
\draw [fill=white] (1.35+11*0.6+0.3,\x*0.8) circle [radius=0.3cm];
\foreach \x in {0,...,9}
\node at (1.35+11*0.6+0.3,\x*0.8) {\scriptsize $Z$};

\end{tikzpicture}
\label{fig:GHZ_circ}
}

\subfloat[][Our largest QFT circuit.]{

\begin{tikzpicture}[scale=0.65, every node/.style={scale=0.85}]

\node at (-1.6,4*0.8) {};

\node at (-1.8,3*0.8) {\footnotesize $\ket{0}_{4}$};
\node at (-1.8,2*0.8) {\footnotesize $\ket{0}_{7}$};
\node at (-1.8,1*0.8) {\footnotesize $\ket{0}_{10}$};
\node at (-1.8,0*0.8) {\footnotesize $\ket{0}_{12}$};

\foreach \x in {0,...,3}
\draw (0.-1.4,\x*0.8) -- (8.55+0.8,\x*0.8);

\foreach \x in {0,1,2,3}
\draw [fill=white] (0.3-1.4,\x*0.8-0.3) rectangle (1.0-1.4,\x*0.8+0.3);
\foreach \x in {0,1,2,3}
\node at (0.65-1.4,\x*0.8) {\scriptsize $H$};

\node at (-1.2,3*0.8+1) {\scriptsize \textcolor{blue}{preparation of}};
\node at (-1.2,3*0.8+0.6) {\scriptsize \textcolor{blue}{$|+\rangle$ states}};
\node at (5,3*0.8+0.8) {\scriptsize \textcolor{blue}{QFT circuit}};
\draw [dashed, thick, blue] (0.,0-0.3) -- (0.,3*0.8+0.7);

\foreach \x in {3}
\draw [fill=white] (0.3,\x*0.8-0.3) rectangle (1.0,\x*0.8+0.3);
\foreach \x in {3}
\node at (0.65,\x*0.8) {\scriptsize $H$};

\draw (1.4,3*0.8) -- (1.4,2*0.8);
\draw [fill=black] (1.4,2*0.8) circle [radius=0.07cm];
\draw [fill=white] (1.4,3*0.8) circle [radius=0.32cm];
\node at (1.4,3*0.8) {\footnotesize $\frac{\pi}{2}$};

\draw (2.35,3*0.8) -- (2.35,1*0.8);
\draw [fill=black] (2.35,1*0.8) circle [radius=0.07cm];
\draw [fill=white] (2.35,3*0.8) circle [radius=0.32cm];
\node at (2.35,3*0.8) {\footnotesize $\frac{\pi}{4}$};

\draw (3.3,3*0.8) -- (3.3,0*0.8);
\draw [fill=black] (3.3,0*0.8) circle [radius=0.07cm];
\draw [fill=white] (3.3,3*0.8) circle [radius=0.32cm];
\node at (3.3,3*0.8) {\footnotesize $\frac{\pi}{8}$};

\foreach \x in {2}
\draw [fill=white] (3.8,\x*0.8-0.3) rectangle (4.5,\x*0.8+0.3);
\foreach \x in {2}
\node at (3.8+0.35,\x*0.8) {\scriptsize $H$};

\draw (4.4+0.7,2*0.8) -- (4.4+0.7,1*0.8);
\draw [fill=black] (4.4+0.7,1*0.8) circle [radius=0.07cm];
\draw [fill=white] (4.4+0.7,2*0.8) circle [radius=0.32cm];
\node at (4.4+0.7,2*0.8) {\footnotesize $\frac{\pi}{2}$};

\draw (5.35+0.7,2*0.8) -- (5.35+0.7,0*0.8);
\draw [fill=black] (5.35+0.7,0*0.8) circle [radius=0.07cm];
\draw [fill=white] (5.35+0.7,2*0.8) circle [radius=0.32cm];
\node at (5.35+0.7,2*0.8) {\footnotesize $\frac{\pi}{4}$};

\foreach \x in {1}
\draw [fill=white] (5.35+1.2,\x*0.8-0.3) rectangle (5.35+1.2+0.7,\x*0.8+0.3);
\foreach \x in {1}
\node at (5.35+1.2+0.35,\x*0.8) {\scriptsize $H$};

\draw (5.35+1.2+0.8+0.5,1*0.8) -- (5.35+1.2+0.8+0.5,0*0.8);
\draw [fill=black] (5.35+1.2+0.8+0.5,0*0.8) circle [radius=0.07cm];
\draw [fill=white] (5.35+1.2+0.8+0.5,1*0.8) circle [radius=0.32cm];
\node at (5.35+1.2+0.8+0.5,1*0.8) {\footnotesize $\frac{\pi}{2}$};

\foreach \x in {0}
\draw [fill=white] (5.35+1.3+1.6,\x*0.8-0.3) rectangle (5.35+1.3+1.6+0.7,\x*0.8+0.3);
\foreach \x in {0}
\node at (5.35+1.3+1.6+0.35,\x*0.8) {\scriptsize $H$};

\foreach \x in {0,...,3}
\draw (8.65+0.8+0.6,\x*0.8-0.05) -- (8.65+0.8,\x*0.8-0.05);
\foreach \x in {0,...,3}
\draw (8.65+0.8+0.6,\x*0.8+0.05) -- (8.65+0.8,\x*0.8+0.05);

\foreach \x in {0,...,3}
\draw [fill=white] (8.65+0.8,\x*0.8) circle [radius=0.3cm];
\foreach \x in {0,...,3}
\node at (8.65+0.8,\x*0.8) {\scriptsize $Z$};

\end{tikzpicture}
\label{fig:QFT_circ}
}

     \subfloat[][Six-qubit depth nine pseudo-random circuit.]{
\begin{tikzpicture}[scale=0.65, every node/.style={scale=0.85}]

\node at (-0.4,6*0.8) {};

\node at (-0.4,5*0.8) {\footnotesize $\ket{0}_{4}$};
\node at (-0.4,4*0.8) {\footnotesize $\ket{0}_{7}$};
\node at (-0.4,3*0.8) {\footnotesize $\ket{0}_{10}$};
\node at (-0.4,2*0.8) {\footnotesize $\ket{0}_{12}$};
\node at (-0.4,1*0.8) {\footnotesize $\ket{0}_{15}$};
\node at (-0.4,0*0.8) {\footnotesize $\ket{0}_{18}$};

\foreach \x in {0,...,5}
\draw (0.,\x*0.8) -- (1.6+4*1.6+0.7,\x*0.8);

\foreach \x in {0,1,2,3,4,5}
\draw [fill=white] (0.3,\x*0.8-0.3) rectangle (1.3,\x*0.8+0.3);
\node at (0.3+0.5,5*0.8) {\scriptsize $V_{4,1}$};
\node at (0.3+0.5,4*0.8) {\scriptsize $V_{7,1}$};
\node at (0.3+0.5,3*0.8) {\scriptsize $V_{10,1}$};
\node at (0.3+0.5,2*0.8) {\scriptsize $V_{12,1}$};
\node at (0.3+0.5,1*0.8) {\scriptsize $V_{15,1}$};
\node at (0.3+0.5,0*0.8) {\scriptsize $V_{18,1}$};

\draw (1.6,4*0.8) -- (1.6,5*0.8);
\draw (1.6,2*0.8) -- (1.6,3*0.8);
\draw (1.6,0*0.8) -- (1.6,1*0.8);
\foreach \x in {0,1,2,3,4,5}
\draw [fill=black] (1.6,\x*0.8) circle [radius=0.07cm];

\foreach \x in {0,1,2,3,4,5}
\draw [fill=white] (0.3+1*1.6,\x*0.8-0.3) rectangle (1.3+1*1.6,\x*0.8+0.3);
\node at (0.3+0.5+1*1.6,5*0.8) {\scriptsize $V_{4,2}$};
\node at (0.3+0.5+1*1.6,4*0.8) {\scriptsize $V_{7,2}$};
\node at (0.3+0.5+1*1.6,3*0.8) {\scriptsize $V_{10,2}$};
\node at (0.3+0.5+1*1.6,2*0.8) {\scriptsize $V_{12,2}$};
\node at (0.3+0.5+1*1.6,1*0.8) {\scriptsize $V_{15,2}$};
\node at (0.3+0.5+1*1.6,0*0.8) {\scriptsize $V_{18,2}$};

\draw (1.6+1*1.6,4*0.8) -- (1.6+1*1.6,3*0.8);
\draw (1.6+1*1.6,1*0.8) -- (1.6+1*1.6,2*0.8);
\foreach \x in {1,2,3,4}
\draw [fill=black] (1.6+1*1.6,\x*0.8) circle [radius=0.07cm];

\foreach \x in {0,1,2,3,4,5}
\draw [fill=white] (0.3+2*1.6,\x*0.8-0.3) rectangle (1.3+2*1.6,\x*0.8+0.3);
\node at (0.3+0.5+2*1.6,5*0.8) {\scriptsize $V_{4,3}$};
\node at (0.3+0.5+2*1.6,4*0.8) {\scriptsize $V_{7,3}$};
\node at (0.3+0.5+2*1.6,3*0.8) {\scriptsize $V_{10,3}$};
\node at (0.3+0.5+2*1.6,2*0.8) {\scriptsize $V_{12,3}$};
\node at (0.3+0.5+2*1.6,1*0.8) {\scriptsize $V_{15,3}$};
\node at (0.3+0.5+2*1.6,0*0.8) {\scriptsize $V_{18,3}$};

\draw (1.6+2*1.6,4*0.8) -- (1.6+2*1.6,5*0.8);
\draw (1.6+2*1.6,2*0.8) -- (1.6+2*1.6,3*0.8);
\draw (1.6+2*1.6,0*0.8) -- (1.6+2*1.6,1*0.8);
\foreach \x in {0,1,2,3,4,5}
\draw [fill=black] (1.6+2*1.6,\x*0.8) circle [radius=0.07cm];

\foreach \x in {0,1,2,3,4,5}
\draw [fill=white] (0.3+3*1.6,\x*0.8-0.3) rectangle (1.3+3*1.6,\x*0.8+0.3);
\node at (0.3+0.5+3*1.6,5*0.8) {\scriptsize $V_{4,4}$};
\node at (0.3+0.5+3*1.6,4*0.8) {\scriptsize $V_{7,4}$};
\node at (0.3+0.5+3*1.6,3*0.8) {\scriptsize $V_{10,4}$};
\node at (0.3+0.5+3*1.6,2*0.8) {\scriptsize $V_{12,4}$};
\node at (0.3+0.5+3*1.6,1*0.8) {\scriptsize $V_{15,4}$};
\node at (0.3+0.5+3*1.6,0*0.8) {\scriptsize $V_{18,4}$};

\draw (1.6+3*1.6,4*0.8) -- (1.6+3*1.6,3*0.8);
\draw (1.6+3*1.6,1*0.8) -- (1.6+3*1.6,2*0.8);
\foreach \x in {1,2,3,4}
\draw [fill=black] (1.6+3*1.6,\x*0.8) circle [radius=0.07cm];

\foreach \x in {0,1,2,3,4,5}
\draw [fill=white] (0.3+4*1.6,\x*0.8-0.3) rectangle (1.3+4*1.6,\x*0.8+0.3);
\node at (0.3+0.5+4*1.6,5*0.8) {\scriptsize $V_{4,5}$};
\node at (0.3+0.5+4*1.6,4*0.8) {\scriptsize $V_{7,5}$};
\node at (0.3+0.5+4*1.6,3*0.8) {\scriptsize $V_{10,5}$};
\node at (0.3+0.5+4*1.6,2*0.8) {\scriptsize $V_{12,5}$};
\node at (0.3+0.5+4*1.6,1*0.8) {\scriptsize $V_{15,5}$};
\node at (0.3+0.5+4*1.6,0*0.8) {\scriptsize $V_{18,5}$};

\foreach \x in {0,...,5}
\draw (1.2+4*1.6+0.7+0.6,\x*0.8-0.05) -- (1.2+4*1.6+0.7,\x*0.8-0.05);
\foreach \x in {0,...,5}
\draw (1.2+4*1.6+0.7+0.6,\x*0.8+0.05) -- (1.2+4*1.6+0.7,\x*0.8+0.05);

\foreach \x in {0,...,5}
\draw [fill=white] (1.2+4*1.6+0.7,\x*0.8) circle [radius=0.3cm];
\foreach \x in {0,...,5}
\node at (1.2+4*1.6+0.7,\x*0.8) {\scriptsize $Z$};

\node at (-1.6,0*0.8-0.5) {};

\end{tikzpicture}
\label{fig:random_circ}
}
\label{fig:}
\caption{\small (a) Our largest GHZ circuit. After adding the QOTP, this circuit contains nine cycles of non-commuting gates. (b) Our largest QFT circuit. We apply the circuit to the state $|+\rangle^{\otimes 4}$. Each two-qubit gate is a controlled-phase gate, which we decompose into two $cZ$ gates interleaved by one-qubit gates \cite{Barron20}. After adding the QOTP, this circuit contains twenty-one cycles of non-commuting gates. (c) Our six-qubit depth nine pseudo-random circuit. The various gates $V_{i,j}$ are random one-qubit gates.}
\end{figure}
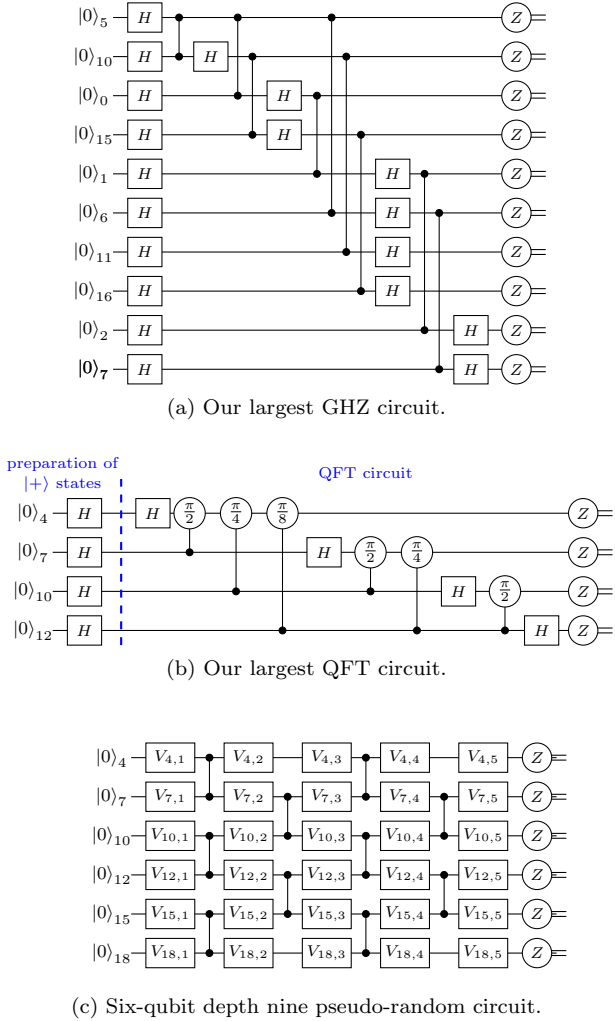

\newpage
\noindent noise that depends \textit{weakly} on the cycles of one-qubit gates (section \hyperlink{sec:gate-dep}2 of the Appendix).

\hypertarget{sec:AP}{\textit{3. Accreditation Protocol|}}Our AP takes as input the target circuit, and two numbers $\theta,\alpha\in(0,1)$ which quantify the desired accuracy and confidence on the final bound.  The target circuit must (i)~take as input~$n$ qubits in the state $\ket{0}$, (ii) contain $2m$ cycles alternating between a cycle of one-qubit gates and a cycle of two-qubit gates and (iii) end with measurements in the Pauli-$Z$ basis (Fig.~\ref{fig:target}). Our AP requires that all two-qubit gates in the target circuit be Clifford gates, so that arbitrary noise processes can be transformed into Pauli errors via QOTP. Without loss of generality, we assume that all the two-qubit gates in the circuit are $cZ$ gates. Note that circuits containing different two-qubit Clifford gates (such as the $cX$ gates implemented by IBM Quantum devices or the M\o{}lmer-S\o{}rensen gate implemented by trapped-ion quantum computers~\cite{MS99}) can be efficiently recompiled in this form without increasing the depth, while circuits containing two-qubit non-Clifford gates (such as those implemented by Google Sycamore~\cite{GoogleSupremacy19}) require a linear increase in depth.

Our AP requires executing $v+1$ circuits sequentially, where $v=\lceil2\textrm{ln}(2/(1-\alpha))/\theta^2\rceil$ and $\lceil\cdot\rceil$ is the ceiling function. One of these circuits (chosen at random) is the target circuit, the others are trap circuits.  Each trap circuit is obtained by replacing the one-qubit gates in the target circuit with random one-qubit Clifford gates as per the following algorithm (Fig. \ref{fig:trap}):
\begin{itemize}[leftmargin=4mm]
\item[1.] For all $j\in\{1,\ldots,m-1\}$ and for all $i\in\{1,\ldots,n\}$:
\begin{itemize}[leftmargin=4mm]
\item[(i)]  If the $j$-th cycle of $cZ$ gates connects qubit $i$ to qubit~$i'$, randomly replace $U_{i,j}$ with $S$ and $U_{i^\prime,j}$ with $H$, or $U_{i,j}$ with $H$ and $U_{i',j}$ with $S$. Undo these gates after the cycle of $cZ$ gates.
\item[(ii)] If the $j$-th cycle of $cZ$ gates does not connect qubit~$i$ to any other qubit, randomly replace $U_{i,j}$ with $H$ or $S$. Undo this gate after the cycle of $cZ$ gates.
\end{itemize}
\item[2.] Initialize a random bit $t\in\{0,1\}$. If $t=0$, do nothing. If $t=1$, append a cycle of Hadamard gates at the beginning and at the end of the circuit. 
\end{itemize}
Since $(S^\dg\otimes H)cZ(S\otimes H)=cX$, the trap circuits apply

\newpage
\onecolumngrid

\begin{figure}[H]
\centering
  \includegraphics[clip,width=.95\columnwidth]{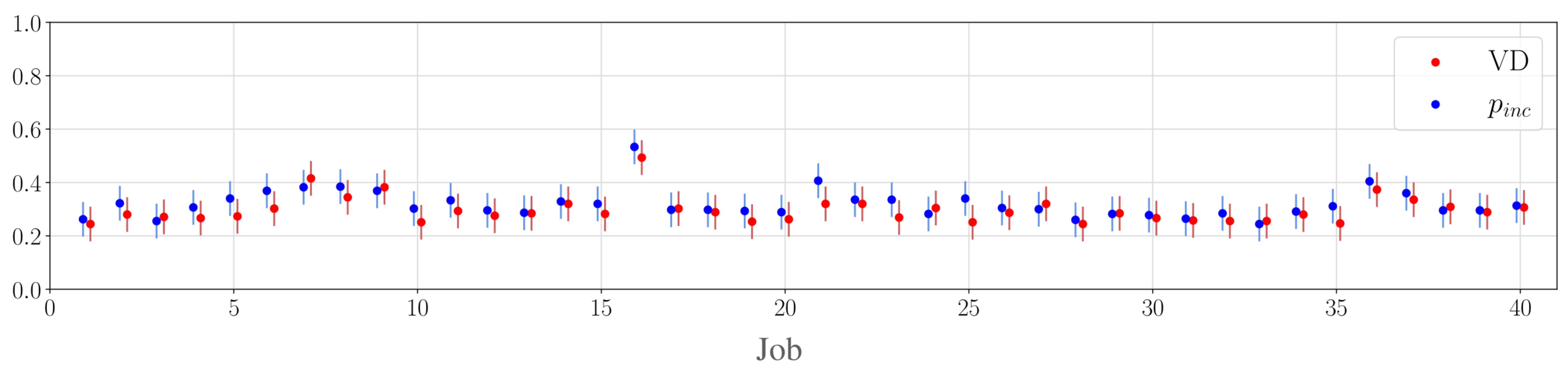}%
\caption{\small Values of VD (red points) and $p_\textrm{inc}$ (blue points) measured in the experiment where the target circuit generates a six-qubit GHZ state. Each job contains 900 circuits (450 target circuits and $450$ independently chosen trap circuits).  The bars correspond to confidence levels above $95\%$|specifically, we set $\theta/2=6.5\%$ and $\alpha\gtrsim 95\%$,  see Eq. \ref{eq:hoeff}.}
\label{fig:GHZ_nq_6}
\end{figure}

\twocolumngrid

\noindent  a series of $cX$ gates to $n$ qubits in the state $\ket{0}$ (if $t=0$) or $\ket{+}$ (if $t=1$). Using $cX\ket{00}=\ket{00}$ and $cX\ket{++}=\ket{++}$ it can be seen that in the absence of noise all the trap circuits return the bit-string $(0,0,\ldots,0)$.

After initializing the $v+1$ circuits,  we append a QOTP to each cycle of one-qubit gates in every circuit.  This is done by appending a cycle of random Pauli gates after every cycle of one-qubit gates, and by appending a second cycle of Pauli gates before the following cycle of one-qubit gates that undoes the first. This randomizes the noise to stochastic Pauli errors \cite{C05,WE16,BFK09,WE16,FKD17,FKD18,Hashim20}.  The trap circuits are designed to \textit{detect} these Pauli errors, meaning that any Pauli error alters their outputs with a probability of at least $50\%$~\cite{FKD18}.

After appending the QOTP we recompile neighbouring cycles of one-qubit gates into a single cycle. This ensures that all the circuits (target and traps) contain the same number of cycles as the circuit given as input to the AP. Next, we implement all the circuits and subsequently estimate the probability $p_\textrm{inc}$ as the fraction $N_\textrm{inc}/v$ of traps that return an incorrect output.  Since $v>2\textrm{ln}(2/(1-\alpha))/\theta^2$, the Hoeffding's inequality~\cite{H63} guarantees that
\begin{equation}
\label{eq:hoeff}
\textrm{prob}\bigg(\bigg|p_\textrm{inc}-\frac{N_\textrm{inc}}{v}\bigg|\leq\frac{\theta}{2}\bigg)\geq\alpha\:.
\end{equation}
Finally, we calculate the bound on the VD as $2N_\textrm{inc}/v$, and we have prob$(|2p_\textrm{inc}-2{N_\textrm{inc}}/{v}|\leq\theta)\geq\alpha$ by Hoeffding's inequality.

The quantity $2p_\textrm{inc}$ grows linearly with the total probability $p_{\textrm{err}}$ that the target circuit is afflicted by errors. More formally, we have
\begin{equation}
\label{eq:p_inc}
    p_\textrm{err}\leq 2p_\textrm{inc}
    \leq 2p_\textrm{err}\:.
\end{equation}
Here, the bound on the l.h.s. is proven in section \hyperlink{sec:proof_bound}1 of the Appendix, while that on the r.h.s. is a consequence of the fact that in the absence of errors the traps always return the correct output. Since the VD is at most unity by construction, it follows that if $p_{\textrm{err}} \leq 50\%$ our AP always returns a \textit{non-trivial} bound on the VD (i.e., below unity). Otherwise, it may return a trivial bound, indicating that the device is afflicted by such high levels of noise that its outputs are far enough from the ideal ones as to be unreliable. 

As can be seen in Fig. \ref{fig:bounds_best_intro}, in our experiments we obtain non-trivial bounds for circuits with up to ten qubits. Larger circuits yield trivial bounds. However, Eq.~\ref{eq:p_inc} shows that improvements in the hardware will extend the reach of the AP beyond ten-qubit circuits. Being fully scalable, in the future the AP will be able to accredit the outputs of quantum circuits that will be intractable for the protocols relying on classical simulations.

\hypertarget{sec:exp}{\textit{4. Experimental Accreditation|}}We implement our AP on two superconducting quantum computers,  $\textsf{ibmq}\_\textsf{johannesburg}$ and $\textsf{ibmq}\_\textsf{paris}$.  These quantum computers consist of superconducting transmon qubits dispersively coupled according to the topology given in Fig.~\ref{fig:ibmq_architecture},  where each edge denotes a $cX$ gate that can be implemented via the cross-resonance interaction.  For a more comprehensive description of this architecture see Ref. \cite{Jurcevic20} and for specific details about $\textsf{ibmq}\_\textsf{johannesburg}$ and $\textsf{ibmq}\_\textsf{paris}$ see Ref.~\cite{IBM}.

We begin by conducting fourteen experiments to accredit the outputs of QFT and GHZ circuits of different widths {(Figs. \ref{fig:GHZ_circ} and \ref{fig:QFT_circ})}. In every experiment we submit 40 jobs to the backend. Each job contains 450 trap circuits, each one chosen independently at random as described in section~\hyperlink{sec:AP}3.  At the end of each job we estimate $p_\textrm{inc}$, as illustrated in Fig.~\ref{fig:GHZ_nq_6} for the preparation and measurement of the six-qubit GHZ state. (See \cite{Git} for more figures).  To demonstrate the AP, in each job we also implement 450 instances of the target circuit and compute the VD between the ideal and experimentally obtained probability distributions. In our experiments this can be done within a reasonable amount of time given the size of the target circuits.

\begin{figure}
     \centering
          \subfloat[][]{

\begin{tikzpicture}[scale=0.65, every node/.style={scale=0.85}]

\foreach \x in {1,...,6}
\node at (-0.3,4.8-\x*0.8) {\footnotesize $\ket{0}_{\x}$};

\foreach \x in {0,...,5}
\draw (0.1,4.0-\x*0.8) -- (5.5,4.0-\x*0.8);

\foreach \x in {0,...,5}
\draw [dashed] (5.5,4.0-\x*0.8) -- (6.5,4.0-\x*0.8);

\foreach \x in {0,...,5}
\draw (11,4.0-\x*0.8) -- (6.5,4.0-\x*0.8);

\foreach \x in {0,...,5}
\draw [fill=white] (0.4,4.0-\x*0.8-0.3) rectangle (1.1,4.0-\x*0.8+0.3);

\foreach \x in {1,...,6}
\node at (0.75,4.8-\x*0.8) {\scriptsize $H^t$};

\foreach \x in {0,...,5}
\draw [fill=white] (1.3,4.0-\x*0.8-0.3) rectangle (2.0,4.0-\x*0.8+0.3);

\node at (1.65,4.8-1*0.8) {\scriptsize $S$};
\node at (1.65,4.8-2*0.8) {\scriptsize $S$};
\node at (1.65,4.8-3*0.8) {\scriptsize $H$};
\node at (1.65,4.8-4*0.8) {\scriptsize $S$};
\node at (1.65,4.8-5*0.8) {\scriptsize $S$};
\node at (1.65,4.8-6*0.8) {\scriptsize $H$};

\draw [fill=black] (2.45,4.8-2*0.8) circle [radius=0.07cm];
\draw [fill=black] (2.45,4.8-3*0.8) circle [radius=0.07cm];
\draw [fill=black] (2.45,4.8-4*0.8) circle [radius=0.07cm];
\draw [fill=black] (2.45,4.8-6*0.8) circle [radius=0.07cm];

\draw (2.45,4.8-2*0.8) -- (2.45,4.8-3*0.8);
\draw (2.45,4.8-4*0.8) -- (2.45,4.8-6*0.8);

\foreach \x in {0,...,5}
\draw [fill=white] (2.8,4.0-\x*0.8-0.3) rectangle (3.6,4.0-\x*0.8+0.3);

\node at (3.2,4.8-1*0.8) {\scriptsize $S^\dagger$};
\node at (3.2,4.8-2*0.8) {\scriptsize $S^\dg$};
\node at (3.2,4.8-3*0.8) {\scriptsize $H$};
\node at (3.2,4.8-4*0.8) {\scriptsize $S^\dagger$};
\node at (3.2,4.8-5*0.8) {\scriptsize $S^\dagger$};
\node at (3.2,4.8-6*0.8) {\scriptsize $H$};

\foreach \x in {0,...,5}
\draw [fill=white] (3.8,4.0-\x*0.8-0.3) rectangle (4.6,4.0-\x*0.8+0.3);

\node at (4.2,4.8-1*0.8) {\scriptsize $H$};
\node at (4.2,4.8-2*0.8) {\scriptsize $S$};
\node at (4.2,4.8-3*0.8) {\scriptsize $S$};
\node at (4.2,4.8-4*0.8) {\scriptsize $S$};
\node at (4.2,4.8-5*0.8) {\scriptsize $S$};
\node at (4.2,4.8-6*0.8) {\scriptsize $H$};

\draw [fill=black] (5.25,4.8-1*0.8) circle [radius=0.07cm];
\draw [fill=black] (4.9+.15,4.8-2*0.8) circle [radius=0.07cm];
\draw [fill=black] (5.25,4.8-5*0.8) circle [radius=0.07cm];
\draw [fill=black] (4.9+.15,4.8-6*0.8) circle [radius=0.07cm];

\draw (4.9+.15,0.0) -- (4.9+.15,3.2);
\draw (5.25,4.0) -- (5.25,0.8);



\foreach \x in {0,...,5}
\draw [fill=white] (6.9,4.0-\x*0.8-0.3) rectangle (7.7,4.0-\x*0.8+0.3);
\foreach \x in {2}
\draw [fill=red!30] (6.9,4.0-\x*0.8-0.3) rectangle (7.7,4.0-\x*0.8+0.3);

\node at (7.3,4.8-1*0.8) {\scriptsize $H$};
\node at (7.3,4.8-2*0.8) {\scriptsize $S$};
\node at (7.3,4.8-3*0.8) {\scriptsize $S$};
\node at (7.3,4.8-4*0.8) {\scriptsize $H$};
\node at (7.3,4.8-5*0.8) {\scriptsize $H$};
\node at (7.3,4.8-6*0.8) {\scriptsize $S$};

\draw [fill=black] (8+.15,4.8-3*0.8) circle [radius=0.07cm];
\draw [fill=black] (8+.15,4.8-4*0.8) circle [radius=0.07cm];

\draw (8+.15,1.6) -- (8+.15,2.4);


\foreach \x in {0,...,5}
\draw [fill=white] (8.6,4.0-\x*0.8-0.3) rectangle (9.4,4.0-\x*0.8+0.3);

\node at (9,4.8-1*0.8) {\scriptsize $H$};
\node at (9,4.8-2*0.8) {\scriptsize $S^\dagger$};
\node at (9,4.8-3*0.8) {\scriptsize $S^\dagger$};
\node at (9,4.8-4*0.8) {\scriptsize $H$};
\node at (9,4.8-5*0.8) {\scriptsize $H$};
\node at (9,4.8-6*0.8) {\scriptsize $S^\dg$};

\foreach \x in {0,...,5}
\draw [fill=white] (9.6,4.0-\x*0.8-0.3) rectangle (10.4,4.0-\x*0.8+0.3);

\foreach \x in {0,...,5}
\node at (10,4.0-\x*0.8) {\scriptsize $H^t$};

\foreach \x in {1,...,6}
\draw (9.1+2.2,4.8-\x*0.8-0.05) -- (9.3+2.2,4.8-\x*0.8-0.05);
\foreach \x in {1,...,6}
\draw (9.1+2.2,4.8-\x*0.8+0.05) -- (9.3+2.2,4.8-\x*0.8+0.05);

\foreach \x in {0,...,5}
\draw [fill=white] (11,\x*0.8) circle [radius=0.3cm];
\foreach \x in {1,...,6}
\node at (9.8+1.2,4.8-\x*0.8) {\scriptsize $Z$};

\end{tikzpicture}
\label{fig:error_future}
     }
     \qquad
     \qquad
     \subfloat[][]{

\begin{tikzpicture}[scale=0.65, every node/.style={scale=0.85}]

\foreach \x in {1,...,6}
\node at (-0.3,4.8-\x*0.8) {\footnotesize $\ket{0}_{\x}$};

\foreach \x in {0,...,5}
\draw (0.1,4.0-\x*0.8) -- (5.5,4.0-\x*0.8);

\foreach \x in {0,...,5}
\draw [dashed] (5.5,4.0-\x*0.8) -- (6.5,4.0-\x*0.8);

\foreach \x in {0,...,5}
\draw (11,4.0-\x*0.8) -- (6.5,4.0-\x*0.8);

\foreach \x in {0,...,5}
\draw [fill=white] (0.4,4.0-\x*0.8-0.3) rectangle (1.1,4.0-\x*0.8+0.3);

\foreach \x in {1,...,6}
\node at (0.75,4.8-\x*0.8) {\scriptsize $H^t$};

\foreach \x in {0,...,5}
\draw [fill=white] (1.3,4.0-\x*0.8-0.3) rectangle (2.0,4.0-\x*0.8+0.3);
\foreach \x in {1}
\draw [fill=red!30] (1.3,4.0-\x*0.8-0.3) rectangle (2.0,4.0-\x*0.8+0.3);

\node at (1.65,4.8-1*0.8) {\scriptsize $S$};
\node at (1.65,4.8-2*0.8) {\scriptsize $S$};
\node at (1.65,4.8-3*0.8) {\scriptsize $H$};
\node at (1.65,4.8-4*0.8) {\scriptsize $S$};
\node at (1.65,4.8-5*0.8) {\scriptsize $S$};
\node at (1.65,4.8-6*0.8) {\scriptsize $H$};

\draw [fill=red!30, draw=red!30] (2.45+0.2,4.0-2*0.8-0.3) rectangle (2.45-0.2,4.0-6*0.8+0.3);

\draw [fill=black] (2.45,4.8-2*0.8) circle [radius=0.07cm];
\draw [fill=black] (2.45,4.8-3*0.8) circle [radius=0.07cm];
\draw [fill=black] (2.45,4.8-4*0.8) circle [radius=0.07cm];
\draw [fill=black] (2.45,4.8-6*0.8) circle [radius=0.07cm];

\draw (2.45,4.8-2*0.8) -- (2.45,4.8-3*0.8);
\draw (2.45,4.8-4*0.8) -- (2.45,4.8-6*0.8);

\foreach \x in {0,...,5}
\draw [fill=white] (2.8,4.0-\x*0.8-0.3) rectangle (3.6,4.0-\x*0.8+0.3);
\foreach \x in {5}
\draw [fill=red!30] (2.8,4.0-\x*0.8-0.3) rectangle (3.6,4.0-\x*0.8+0.3);

\node at (3.2,4.8-1*0.8) {\scriptsize $S^\dagger$};
\node at (3.2,4.8-2*0.8) {\scriptsize $S^\dg$};
\node at (3.2,4.8-3*0.8) {\scriptsize $H$};
\node at (3.2,4.8-4*0.8) {\scriptsize $S^\dagger$};
\node at (3.2,4.8-5*0.8) {\scriptsize $S^\dagger$};
\node at (3.2,4.8-6*0.8) {\scriptsize $H$};

\foreach \x in {0,...,5}
\draw [fill=white] (3.8,4.0-\x*0.8-0.3) rectangle (4.6,4.0-\x*0.8+0.3);

\node at (4.2,4.8-1*0.8) {\scriptsize $H$};
\node at (4.2,4.8-2*0.8) {\scriptsize $S$};
\node at (4.2,4.8-3*0.8) {\scriptsize $S$};
\node at (4.2,4.8-4*0.8) {\scriptsize $S$};
\node at (4.2,4.8-5*0.8) {\scriptsize $S$};
\node at (4.2,4.8-6*0.8) {\scriptsize $H$};

\draw [fill=black] (5.25,4.8-1*0.8) circle [radius=0.07cm];
\draw [fill=black] (4.9+.15,4.8-2*0.8) circle [radius=0.07cm];
\draw [fill=black] (5.25,4.8-5*0.8) circle [radius=0.07cm];
\draw [fill=black] (4.9+.15,4.8-6*0.8) circle [radius=0.07cm];

\draw (4.9+.15,0.0) -- (4.9+.15,3.2);
\draw (5.25,4.0) -- (5.25,0.8);

\foreach \x in {0,...,5}
\draw [fill=white] (6.9,4.0-\x*0.8-0.3) rectangle (7.7,4.0-\x*0.8+0.3);
\foreach \x in {2}
\draw [fill=red!30] (6.9,4.0-\x*0.8-0.3) rectangle (7.7,4.0-\x*0.8+0.3);

\node at (7.3,4.8-1*0.8) {\scriptsize $H$};
\node at (7.3,4.8-2*0.8) {\scriptsize $S$};
\node at (7.3,4.8-3*0.8) {\scriptsize $S$};
\node at (7.3,4.8-4*0.8) {\scriptsize $H$};
\node at (7.3,4.8-5*0.8) {\scriptsize $H$};
\node at (7.3,4.8-6*0.8) {\scriptsize $S$};

\draw [fill=red!30, draw=red!30] (8+.15+0.2,4.0-1*0.8-0.3) rectangle (8+.15-0.2,4.0-4*0.8+0.3);

\draw [fill=black] (8+.15,4.8-3*0.8) circle [radius=0.07cm];
\draw [fill=black] (8+.15,4.8-4*0.8) circle [radius=0.07cm];

\draw (8+.15,1.6) -- (8+.15,2.4);

\foreach \x in {0,...,5}
\draw [fill=white] (8.6,4.0-\x*0.8-0.3) rectangle (9.4,4.0-\x*0.8+0.3);

\node at (9,4.8-1*0.8) {\scriptsize $H$};
\node at (9,4.8-2*0.8) {\scriptsize $S^\dagger$};
\node at (9,4.8-3*0.8) {\scriptsize $S^\dagger$};
\node at (9,4.8-4*0.8) {\scriptsize $H$};
\node at (9,4.8-5*0.8) {\scriptsize $H$};
\node at (9,4.8-6*0.8) {\scriptsize $S^\dg$};

\foreach \x in {0,...,5}
\draw [fill=white] (9.6,4.0-\x*0.8-0.3) rectangle (10.4,4.0-\x*0.8+0.3);

\foreach \x in {0,...,5}
\node at (10,4.0-\x*0.8) {\scriptsize $H^t$};

\foreach \x in {1,...,6}
\draw (9.1+2.2,4.8-\x*0.8-0.05) -- (9.3+2.2,4.8-\x*0.8-0.05);
\foreach \x in {1,...,6}
\draw (9.1+2.2,4.8-\x*0.8+0.05) -- (9.3+2.2,4.8-\x*0.8+0.05);

\foreach \x in {0,...,5}
\draw [fill=white] (11,\x*0.8) circle [radius=0.3cm];
\draw [fill=red!30] (11,3*0.8) circle [radius=0.3cm];
\draw [fill=red!30] (11,5*0.8) circle [radius=0.3cm];
\foreach \x in {1,...,6}
\node at (9.8+1.2,4.8-\x*0.8) {\scriptsize $Z$};

\end{tikzpicture}
\label{fig:error_today}
     }

     \caption{\small Examples of trap circuits affected by errors (the faulty gates and measurements are highlighted in red). (a) Single-cycle pattern affecting a single qubit. Patterns of this type are detected with probability 50$\%$ (see Ref.~\cite{FKD18}).  (b) Multi-cycle pattern. Patterns of this type are detected with probability greater than 50$\%$.  }
     \label{fig:errors_example}
\end{figure}
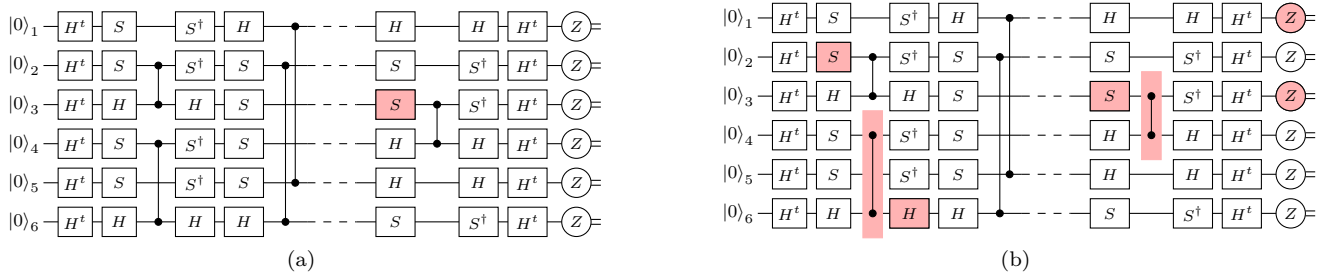

In every job we find VD $\approx$ $p_{\textrm{inc}}$, but we observe fluctuations across different jobs. These fluctuations indicate that different jobs suffer different noise due to e.g. automatic recalibration of the internal components of the device. VD $\approx$ $p_{\textrm{inc}}$ also suggests that the factor 2 on the r.h.s. of Eq.  \ref{eq:boundVD} may be unnecessary. However, this factor~2 captures the effects of specific patterns of errors that are detected with probability 50$\%$ (such as single-cycle patterns afflicting a single qubit, see Fig. \ref{fig:error_future}). Thus, it

\newpage

\onecolumngrid

\twocolumngrid

\noindent is necessary to ensure the validity of Eq.  \ref{eq:boundVD} for arbitrary types of noise. In general, multi-cycle patterns (Fig. \ref{fig:error_today}) as well as patterns afflicting more than one qubit (such as those afflicting the today's devices~\cite{HFW19, Hashim20}) are detected with probability greater than $50\%$ \cite{FKD18}, and for these patterns we have $p_{\textrm{err}}<2p_{\textrm{inc}}$.

Importantly, we find VD$\:<2p_\textrm{inc}$ for every job in each of our experiments \cite{Git}. This proves that in all the tests that we have conducted, our AP has correctly bounded the VD as expected from Eq. ~\ref{eq:boundVD}. Fig.~\ref{fig:bounds_best_intro}\textcolor{brown}{a} shows the smallest value of $2p_{\textrm{inc}}$ obtained in the various experiments.  

To study how $2p_{\textrm{inc}}$ varies with the circuit depth we conduct ten more experiments on $\textsf{ibmq}\_\textsf{paris}$. We target a set of six-qubit pseudo-random circuits of depths ranging from one to nineteen. These circuits alternate cycles of random one-qubit gates to cycles containing either two or three $cZ$ gates {(Fig. \ref{fig:random_circ})}. In every experiment we submit 20 jobs to the backend, each one containing 900 unique trap circuits.  In Fig. \ref{fig:bounds_best_intro}\textcolor{brown}{b} we show the smallest values of $2p_{\textrm{inc}}$ obtained across the 20 jobs for each experiment. 

\hypertarget{sec:diagnostic}{\textit{5. Hardware diagnosis using AP|}}The trap circuits implement deterministic computations,  designed to return the output $\overline{s}=(0,\ldots,0)$ in the absence of errors and some other output in the presence of errors. Importantly, different errors alter the traps' outputs in different ways. Therefore, we expect that the probability distribution of the traps' outputs contains information regarding the nature of the noise afflicting the device in use. To corroborate this,  in this section we focus on the traps' outputs collected in the experiments with six-qubit pseudo-random circuits. We show how these outputs can help identify the main sources of errors in circuits of different sizes implemented on $\textsf{ibmq}\_\textsf{paris}$. 

In $\textsf{ibmq}\_\textsf{paris}$ the error rates provided by the backend are around $0.05\%$ for the one-qubit gates,  $1.5\%$ for the two-qubit gates and $2.3\%$ for single-qubit measurements~\cite{Git}, while errors in state preparation are expected to be negligible. Therefore, we expect measurement noise to be the dominant source of error in shallow circuits and gate noise in deep circuits.  To verify this, let us consider a noise model where the gates are noiseless, while measurement errors flip each bit $s_i\in\overline{s}$ with probability $p_{\textrm{flip}}$. In this scenario, the probability that a trap returns an output $\overline{s}$ with Hamming weight $H_{\overline{s}}=h\in\{0,\ldots,n\}$ is
\begin{equation}
\label{eq:hamming}
P_{\textrm{trap}}\big(H_{\overline{s}}=h\big)=\binom{n}{h}p_{\textrm{flip}}^{h}(1-p_{\textrm{flip}})^{n-h}\:,
\end{equation}
where the Hamming weight $H_{\overline{s}}=\sum_{s_i\in\overline{s}}s_i$ is the number of bits equal to 1 in the output string $\overline{s}$.

\begin{figure}[H]
\flushleft
\subfloat[][]{\includegraphics[clip,width=0.7\columnwidth]{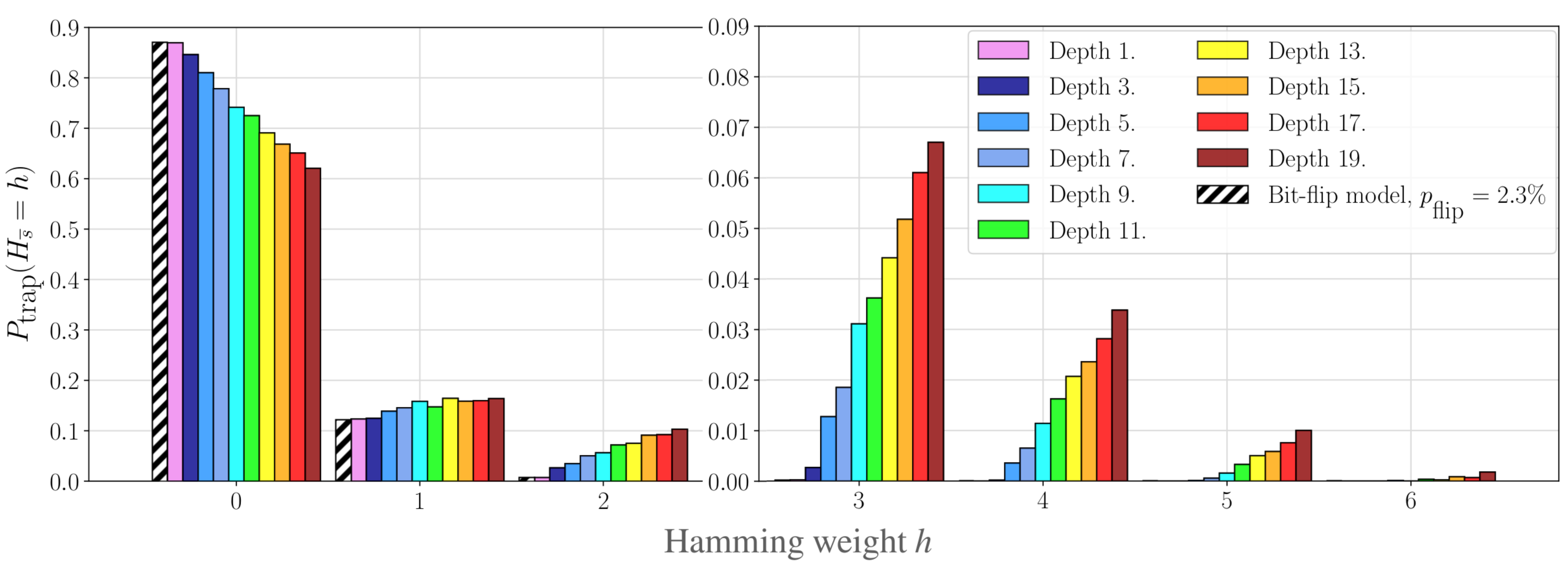}
\label{fig:hamming_all}}
     \subfloat[][]{\includegraphics[clip,width=0.305\columnwidth]{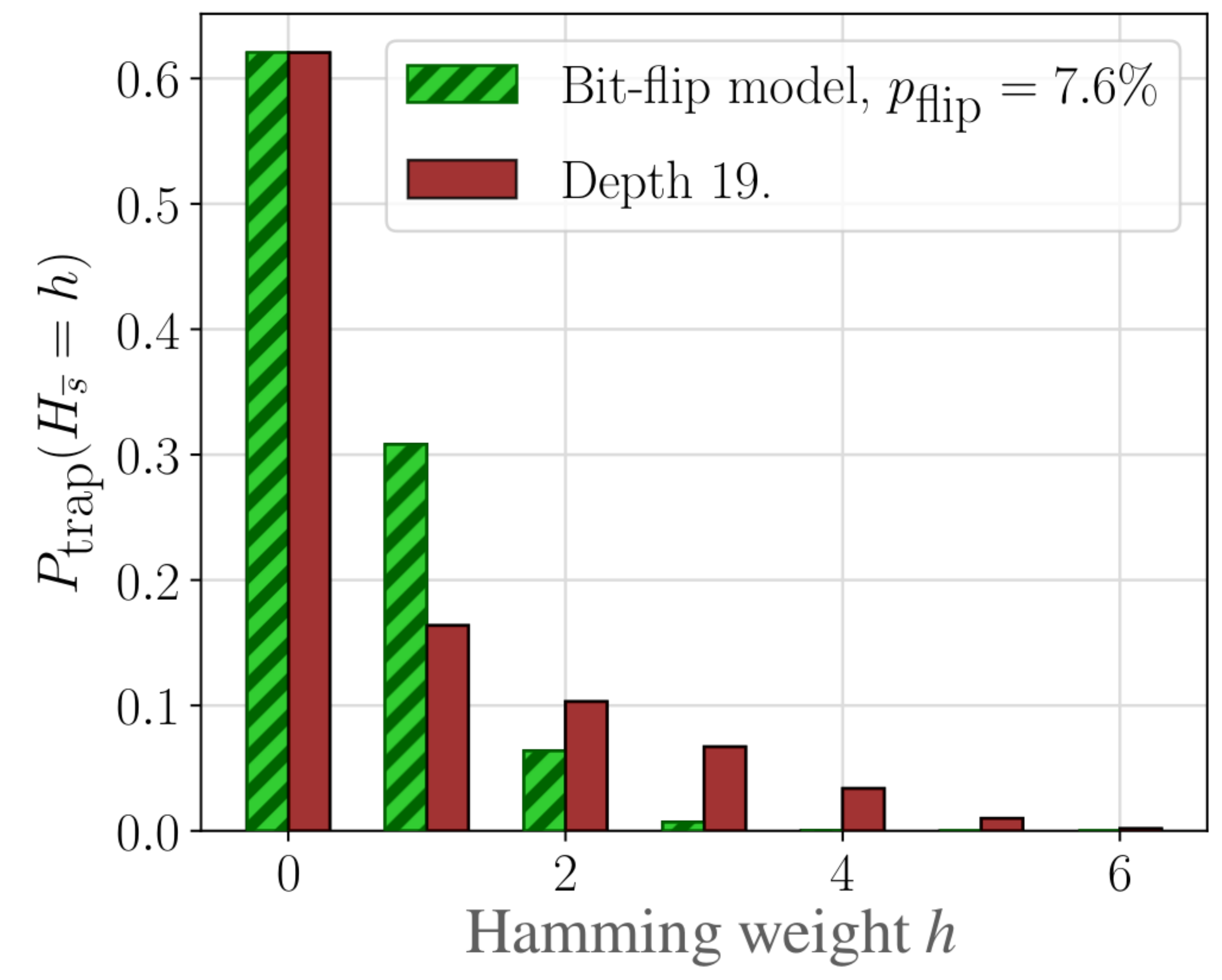}
     \label{fig:hamming_depth19}}
     \caption{The values of $P_{\textrm{trap}}(H_{\overline{s}}=h)$ calculated using the bit-flip model defined in Eq. \ref{eq:hamming} (striped bars) and those measured in the experiments with pseudo-random circuits (solid bars)|for $h\geq3$ the $y$ axis is rescaled.  At any given depth, the values of $P_{\textrm{trap}}(H_{\overline{s}}=h)$ are measured using the outputs of 17980 traps.  In \textbf{(a)} we set $p_{\textrm{flip}}=2.3\%$,  which coincides with the error rate provided for the measurements by the backend.  In \textbf{(b)} we set $p_{\textrm{flip}}=7.6\%$.
     }
     \label{fig:hamming}
\end{figure}

Setting $p_{\textrm{flip}}=2.3\%$, in Fig.~\ref{fig:hamming_all} we compare the values of $P_{\textrm{trap}}\big(H_{\overline{s}}=h\big)$ from our bit-flip noise model (striped bars) with the experimentally  measured  ones  for pseudo-random circuits (solid bars).  It can be seen that the bit-flip model accurately predicts the results obtained for shallow circuits (e.g.  for circuits of depth one or three), indicating that measurement noise dominates short-depth circuits.  It can also be seen that the bit-flip model becomes progressively disparate as the depth increases, indicating that in deep circuits measurements are no more the dominant contributor to noise.

The  above  inference  may  be  challenged  by  positing that  the  measurement  noise  changes  with  the  circuit’s depth.  To rule this possibility out,  in Fig. \ref{fig:hamming_depth19} we set $p_{\textrm{flip}}=7.6\%$ such that the value of $P_{\textrm{trap}}\big(H_{\overline{s}}=0\big)$ calculated using the bit-flip model (left-most bar in the figure) equals the value measured in the experiment with depth nineteen random circuits.  As can be seen in the figure,  the bit-flip model still remains largely disparate.  Overall, measurement errors alone cannot explain the distribution of outputs of our deepest trap circuits and gate noise can no longer be neglected.

This simple analysis builds upon the error rates provided by the backend and is thus device-specific. It shows that the probability distribution of the traps' outputs contains information regarding the noise afflicting $\textsf{ibmq}\_\textsf{paris}$.  Obvious questions as to how much of this information can be retrieved, and whether it can be retrieved in  a  device-agnostic manner  remain  open for future work.

\textit{5. Conclusions|}We have presented an accreditation protocol that uses random Clifford circuits to ascertain the correctness of the outputs of quantum computations implemented on existing hardware.  We have experimen-

\newpage
\onecolumngrid

\twocolumngrid

\noindent tally  demonstrated  its  present  practicality  and  mathematically established its future scalability.

Presently,  the factor 2 in the r.h.s. of Eq.  \ref{eq:boundVD} represents the main obstacle towards increasing the number of qubits in our experiments beyond $n=10$. Indeed, for target circuits with $n>10$ qubits we find $p_{\textrm{inc}}>50\%$, hence our AP returns a  bound on the VD that exceeds unity. This is a trivial bound, since the VD is below unity by construction~\cite{NC00}. Nevertheless, better devices will extend the reach of our AP beyond 10-qubit circuits. Being fully scalable, we anticipate that in the future our AP will replace the protocols based on classical simulations of quantum circuits \cite{B&al16,GoogleSupremacy19,CBSNG19} and will become a standard routine to characterize the outputs of noisy quantum computers.

\textit{Acknowledgments|}SF  and  AD  were  supported  by the  UK  Networked  Quantum  Information  Technologies~(NQIT) Hub (EP/M013243/1) in the early stages of this work. SM and DM research was sponsored by the Army Research Office and was accomplished under Grant Numbers W911NF-14-1-0124 and W911NF-21-1-0002. The views and conclusions contained in this document are those of the authors and should not be interpreted as representing the official policies, either expressed or implied, of the Army Research Office or the U.S. Government. The U.S. Government is authorized to reproduce and distribute reprints for Government purposes notwithstanding any copyright notation herein.

\bibliographystyle{unsrt}
\bibliography{biblio}

\onecolumngrid
\newpage
\twocolumngrid

\section*{Appendix}
The Appendix is organized as follows: In section \hyperlink{sec:proof_bound}1 we provide a derivation of the bound on the VD provided by our AP, in section \hyperlink{sec:gate-dep}2 we show that our protocol is robust to noise processes that depend weakly on the choice of one-qubit gates, in section \hyperlink{sec:comparison}3
we compare our AP with the AP in Ref. \cite{FKD18}. We refer the reader to section~\hyperlink{sec:notation}2 of the main text for the notation.

\hypertarget{sec:proof_bound}{\textit{1. Derivation of the bound on the VD|}}In this section we derive the bound on the VD provided by our AP (Eq.  \ref{eq:boundVD}).  Before presenting the mathematical proof, we calculate the state of the system at the end of a noisy implementation of the $k$-th circuit executed in our AP, with $k\in\{1,\ldots,v+1\}$. 

Under the assumptions that noise is Markovian and that the cycles of one-qubit gates suffer gate-independent noise, the state of the system at the end of a noisy implementation of circuit $k$ is
\begin{align}
\label{eq:before-twirl}
\widetilde{\rho}^{\:(k)}_{\textrm{out}}=&\;{\M}\;\E_m\U^{(k)}_m\:\E_{c\Z_{m-1},m-1}\:c\Z_{m-1}\U_{m-1}^{(k)}\cdots\cr
&\cdots \E_{c\Z_{1},1}\;c\Z_{1}\U^{(k)}_1\;\R\big(|0\rangle\langle0|^{\otimes n}\big)\;,
\end{align}
where $\R$ is the noise in state preparation, $\U_j^{(k)}$ ($c\Z_j$) is the $j$-th cycle of one-qubit gates (two-qubit gates), $\E_{c\Z_{j}, j}$ is the noise due to $c\Z_j\U_j^{(k)}$ (which depends only on $c\Z_j$ and not on $\U_j^{(k)}$) and finally, $\M$ is the round of measurements. To simplify the structure of the noise, a QOTP is appended to each cycle of one-qubit gates in all the circuits. This randomizes the noise into stochastic Pauli errors \cite{WE16,C05,BFK09,FKD17,FKD18} and allows rewriting $\widetilde{\rho}^{\:(k)}_{\textrm{out}}$ as
\begin{align}
\label{eq:end_k_circ}
\widetilde{\rho}^{\:(k)}_{\textrm{out}}=&\sum_{\smash{\p_0,\ldots,\p_m}}q_0(\p_0)\cdots q_m(\p_m)\M\p_mc\Z_m\U_m^{(k)}\cdots\nonumber\\
&\cdots\p_1c\Z_1\U_1^{(k)}\p_0\big(|0\rangle\langle0|^{\otimes n}\big)\:,
\end{align}
where $q_0(\p_0)\cdots q_m(\p_m)$ is the probability that the ``pattern of Pauli errors'' $\p_0,\ldots,\p_m\in\{\I,\X,\Y,\Z\}^{\otimes n}$ occurs. 

Importantly, note that the cycles of two-qubit gates are identical in all the circuits (target and traps), as well as the input state and measurements. Therefore, under the assumption that the one-qubit gates suffer gate-independent noise, the probabilities $q(\p_0)\cdots q(\p_m)$ are the same in all the circuits, and so is the total probability of error per circuit
\begin{equation}
\label{eq:p_err}
p_{\textrm{err}}\;={\sum_{\p_0,\ldots,\p_m\neq\I,\ldots,\I}}q_0(\p_0)\cdots q_m(\p_m)\:.
\end{equation}
We can now establish the bound on the VD.

\begin{proof}\textit{(Eq.  \ref{eq:boundVD}).}
To prove the inequality we make use of the following two statements:\\

\noindent\hypertarget{st1}{\textbf{Statement 1. }}\textit{(Proof in Appendix B of Ref. \cite{FKD18}). Suppose that a trap circuit is afflicted by a ``single-cycle'' pattern of errors, i.e. a pattern such that $\p_{j_0}\neq\I$ for some $j_0\in\{0,\ldots,m\}$ and $\p_j=\I$ for all $j\neq j_0$. Then, summed over the random one-qubit gates in the trap, the trap returns an incorrect output with probability $50\%$ or above.}\\

\noindent\hypertarget{st2}{\textbf{Statement 2. }}\textit{(Proof at the end of this section). Let $q_{\textup{tot}}(j)=\sum_{\p_j\neq\I}q_j(\p_j)$ be the error rate of cycle $j\in\{0,\ldots,m\}$. Denoting by $p_\textup{canc}$ the probability that errors in different cycles of a trap circuit cancel with each other, we have $p_\textup{canc}\leq C$, where
\begin{equation}
\label{eq:C}
    C=O\bigg(\sum_{j,j'\neq j}q_{\textup{tot}}(j)q_{\textup{tot}}(j')\bigg)\:.
\end{equation}
}
Statements \hyperlink{st1}1 and \hyperlink{st2}2 ensure that the trap circuits can detect errors with probability greater that $50\%$. To see this, consider the state of the system at the end of a trap circuit (Eq. \ref{eq:end_k_circ}). Since all the gates in the trap circuits are Clifford, we can map arbitrary patterns of errors into single-cycle patterns. That is, we can commute the errors with the various cycles and \textit{merge} them into a single error $\Q_{(\p_0,\ldots,\p_m)}$ (which depends on the initial errors $\p_0,\ldots,\p_m$), obtaining
\begin{align}
\widetilde{\rho}^{\:(k)}_{\textrm{out}}&=\sum_{\smash{\p_0,\ldots,\p_m}}q_0(\p_0)\cdots q_m(\p_m)\:\M c\Z_m\U_m^{(k)}\cdots\\
&\ldots\Q_{(\p_0,\cdots,\p_m)}c\Z_{j_0}\U_{j_0}^{(k)}\ldots c\Z_1\U_1^{(k)}\big(|0\rangle\langle0|^{\otimes n}\big)\:\nonumber
\end{align}
for some $j_0\in\{0,\ldots,m\}$. In principle, the errors in the trap may cancel with each other, yielding $\Q_{(\p_0,\ldots,\p_m)}=\I$. In particular, denoting by $p_{\textup{canc}}$ the probability of error cancellation, we obtain $\Q_{(\p_0,\ldots,\p_m)}\neq\I$ with probability $p_{\textup{err}}(1-p_{\textup{canc}})$. 

Having mapped the original pattern into a single-cycle pattern, Statement \hyperlink{st1}1 ensures that if errors do not cancel (i.e., if $\Q_{(\p_0,\ldots,\p_m)}\neq\I$), the trap circuit returns the incorrect output with probability greater than $50\%$. This proves that
\begin{equation}
\label{eq:bound_p_inc}
p_{\textrm{inc}}\geq \frac{p_{\textrm{err}}(1-p_{\textrm{canc}})}{2}\:,
\end{equation}
where $p_{\textrm{inc}}$ is the probability that a trap returns an incorrect output.

We can now use Eq.  \ref{eq:bound_p_inc} to upper-bound the VD between ideal and experimental outputs of the target circuit. Labeling the target circuit with $v_0\in\{1,\ldots,v+1\}$, we rewrite the state of the system at the end of the target circuit (Eq. \ref{eq:end_k_circ} with $k=v_0$) as
\begin{equation}
\widetilde{\rho}_{\textup{out}}^{\:(v_0)}=(1-p_\textup{err})\rho_{\textup{out}}^{(v_0)}+p_\textup{err}\sigma^{(v_0)}\:,
\end{equation}
where ${\rho}_{\textup{out}}^{\:(v_0)}$ is the state of the system at the end of an ideal implementation of the target circuit and $\sigma^{(v_0)}$ is a state encompassing the effects of noise. This leads to
\begin{align}
\textrm{VD} =&\frac{1}{2}\sum_{\overline{s}}{\big|p_{\textup{ideal}}(\overline{s})-p_{\textup{exp}}(\overline{s})\big|}\\ =&D\big({\rho}_{\textup{out}}^{\:(v_0)},\widetilde{\rho}_{\textup{out}}^{\:(v_0)}\big)
\leq p_{\textrm{err}}\leq 2 \frac{p_{\textrm{inc}}}{1-p_{\textrm{canc}}}\:,
\end{align}
where $D(\tau,\:\tau')=$Tr$|\tau-\tau'|/2$ is the trace distance between the states $\tau$ and $\tau'$. Finally, since  $p_{\textrm{canc}}\leq C$ and $C$ is quadratic in the cycles' error rates by Statement \hyperlink{st2}2, we have $p_{\textrm{canc}}\ll1$ and
\begin{equation}
\textrm{VD} \leq 2p_{\textrm{inc}}(1+p_{\textrm{canc}})\approx
2p_{\textrm{inc}}
\:.
\end{equation}
\end{proof}

Relying on Statement \hyperlink{st2}2, in the proof of Eq. \ref{eq:boundVD} we used $p_{\textrm{canc}}\leq C$, as well as $C\ll1$. The latter can be corroborated empirically using calibration data. For example, our largest circuit (the ten-qubit GHZ circuit, Fig. \ref{fig:GHZ_circ}) contains five cycles of one-qubit gates with an error rate $\approx0.1\%$ \cite{Git} and four cycles of two-qubit gates. Since each two-qubit gate has an error rate $\approx1.5\%$ \cite{Git}, we estimate an error rate $\approx1.5\%$ for the first cycle, $\approx3\%$ for the second and the fourth and $\approx6\%$ for the third. One-qubit measurements have error rates $\approx2\%$ \cite{Git}, from which we estimate an error rate $\approx20\%$ for the final cycle of measurements. Overall, using Eq. \ref{eq:C} we estimate $C\approx3\%$. With the same strategy we estimate values of $C$ below $3\%$ for all the other circuits. As we point out at the end of this section, $p_{\textrm{canc}}\leq C$ is a loose bound and we expect that $p_{\textrm{canc}}$ be well below $C$ in practice.

We now provide a proof of Statement \hyperlink{st2}2.

\begin{proof}
\textit{(Statement \hyperlink{st2}2.) }For simplicity, let us first consider the case where errors afflict two neighbouring cycles $j$ and $j+1$ and no other cycle. In this case, error cancellation happens when $\p_{j+1}=c\Z_{j+1}\U_{j+1}(\p_{j})$. Therefore, indicating by $Q(j,j+1)$ the probability of error cancellation we have 
\begin{align}
    Q(j,j+1)=&\sum_{\p_j\neq\I}q_j\big(\p_j\big)q_{j+1}\big(c\Z_{j+1}\U_{j+1}(\p_{j})\big)\label{eq:loose1}\\
    \leq&\sum_{\p_j,\p_{j+1}\neq\I}q_j\big(\p_j\big)q_{j+1}\big(\p_{j+1}\big)\label{eq:loose2}\\
    =&\:q_{\textrm{tot}}(j)q_{\textrm{tot}}(j+1)\:,
\end{align}
where to obtain Eq. \ref{eq:loose2} we use the fact that the probability of error cancellation is no more than the product of the probabilities of errors happening. With the same arguments we can upper-bound the probability $Q(j_1,j_2)$ of error cancellation for patterns afflicting any two cycles $j_1$ and $j_2$ as
\begin{align}
    Q(j_1,j_2)
    \leq&q_{\textrm{tot}}(j_1)q_{\textrm{tot}}(j_2)\:.
\end{align}
This proves that the probability of error cancellation for patterns afflicting two cycles is at most quadratic in the cycles' error rates. 

With the same strategy it can be shown that the probability of error cancellation for patterns afflicting $K>2$ cycles $j_1,j_2,\ldots,j_K$ is higher order in the cycles' error rates. Specifically, indicating this probability by $Q(j_1,j_2,\ldots,j_K)$ we find
\begin{align}
    Q(j_1,j_2,\ldots,j_K)
    \leq&\:q_{\textrm{tot}}(j_1)q_{\textrm{tot}}(j_2)\cdots q_{\textrm{tot}}(j_K).
\end{align}
This leads to
\begin{align}
    p_{\textup{canc}}=&\sum_{K\in\{2,\ldots,m+1\}}\bigg(\sum_{j_1,j_2\ldots,j_K}Q(j_1,j_2\ldots,j_K)\bigg)\\
    =&O\bigg(\sum_{j_1,j_2\neq j_1}Q(j_1,j_2)\bigg)\label{eq:Q12}\\
    =&O\bigg(\sum_{j_1,j_2\neq j_1}q_{\textrm{tot}}(j_1)q_{\textrm{tot}}(j_2)\bigg)\:.
\end{align}
\end{proof}
We conclude the section by pointing out that $Q({j_1},j_2,\ldots,j_K)\leq q_{\textrm{tot}}(j_1)q_{\textrm{tot}}(j_2)\cdots q_{\textrm{tot}}(j_K)$, and consequently $p_{\textrm{canc}}\leq C$, is a loose bound. To see this, note that to upper-bound the r.h.s. of Eq.~\ref{eq:loose1} we use $q_{j+1}\big(c\Z_{j+1}\U_{j+1}(\p_{j})\big)\leq q_{\textrm{tot}}(j+1)$. That is, we replace the probabilities of individual errors (including negligible probabilities) with the total probability of error in cycle $j+1$. Based on this observation, we expect that $p_{\textrm{canc}}$ be well below $C$.

\hypertarget{sec:gate-dep}{\textit{2. Robustness to weak gate-dependent noise.}} In the proof of Eq. \ref{eq:boundVD} we have assumed that the cycles of one-qubit gates suffer gate-independent noise. In practice this assumption may be too stringent. To relax this assumption, in this section we analyse how gate-dependent noise may affect the effectiveness of our AP. Formally:
\begin{theorem}
\label{th:gate-dep-noise}
Let us consider a circuit implementing the operation
\begin{equation}
    \C=\sum_{\U_1,\ldots,\U_m}p(\U_1,\ldots,\U_m)c\Z_m\U_m\ldots c\Z_1\U_1\:,
\end{equation}
where the cycles of one-qubit gates $\U_1,\ldots,\U_m$ are chosen with probability $p(\U_1,\ldots,\U_m)$. Let
\begin{align}
    &\C_{\textrm{gi}}=\\&\sum_{\U_1,\ldots,\U_m}p(\U_1,\ldots,\U_m)\E_{c\Z_m,m}c\Z_m\U_m\ldots \E_{c\Z_1,1}c\Z_1\U_1\nonumber
\end{align}
be a noisy implementation of $\C$ with noise $\E_{c\Z_j,j}$ that depends only on the cycle of two-qubit gates $c\Z_j$ and on the index $j$. Let 
\begin{align}
    &\C_{\textrm{gd}}=\\&\sum_{\U_1,\ldots,\U_m}p(\U_1,\ldots,\U_m)\E_{c\Z_m\U_m,m}c\Z_m\U_m\ldots \E_{c\Z_1\U_1,1}c\Z_1\U_1\nonumber
\end{align}
be a noisy implementation of $\C$ with noise $\E_{c\Z_j\U_j,j}$ that depends also on the cycle of one-qubit gates $\U_j$. Averaged over all possible choices of one-qubit gates we have
\begin{equation}
    ||\C_{\textrm{gi}}-\C_{\textrm{gd}}||_\diamond \leq \sum_{\substack{\U_1,\ldots,\U_m\\j=1,\ldots,m}}p(\U_1,\ldots,\U_m)\:||\E_{c\Z_j,j}-\E_{c\Z_j\U_j,j}||_\diamond\:,
\end{equation}
where $||\cdot||_\diamond$ is the diamond distance.
\end{theorem}
The above theorem shows that if the noise depends \textit{weakly} on the choice of one-qubit gates (i.e., $||\E_{c\Z_j,j}-\E_{c\Z_j\U_j,j}||_\diamond$ is small for all $j$), the outputs of a circuit affected by gate-dependent noise remain close to those of the same circuit affected by gate-independent noise. This theorem is valid for any circuit where the one-qubit gates are selected at random. Applied to the target and trap circuits discussed in this paper, it guarantees our AP is robust to noise that depends weakly on the choice of one-qubit gates.

\begin{proof}
\textit{(Theorem \ref{th:gate-dep-noise}).} Our proof follows the same arguments as those in Ref. \cite{WE16}. Let 
\begin{align}
    \F_j&=\E_{c\Z_j,j}c\Z_j\U_j\\
    \G_j&=\E_{c\Z_j\U_j,j}c\Z_j\U_j
\end{align}
and
\begin{align}
    \F_{j:1}&=\F_j\cdots\F_1\\
    \G_{j:1}&=\G_j\cdots\G_1\:.
\end{align}
By induction it can be proven that
\begin{align}
    \F_{m:1}-\G_{m:1}=\sum_{j=1}^m
    \F_{m:j+1}(\F_j-\G_j)\G_{j-1:1}
\end{align}
Noting that
\begin{align}
    \C_{\textrm{gi}}&=\sum_{\U_1,\ldots,\U_m}p(\U_1,\ldots,\U_m)\F_{m:1}\\
    \C_{\textrm{gd}}&=\sum_{\U_1,\ldots,\U_m}p(\U_1,\ldots,\U_m)\G_{m:1}
\end{align}
we have

\begin{align}
    &||\C_{\textrm{gi}}-\C_{\textrm{gd}}||_\diamond \\
    = & \bigg|\bigg|\sum_{\substack{\U_1,\ldots,\U_m\\j=1,\ldots,m}}p(\U_1,\ldots,\U_m)\F_{m:j+1}(\F_j-\G_j)\G_{j-1:1}\bigg|\bigg|_\diamond\\
    \leq & \sum_{\substack{\U_1,\ldots,\U_m\\j=1,\ldots,m}}p(\U_1,\ldots,\U_m)\big|\big|\F_{m:j+1}(\F_j-\G_j)\G_{j-1:1}\big|\big|_\diamond \\
    \leq & \sum_{\substack{\U_1,\ldots,\U_m\\j=1,\ldots,m}}p(\U_1,\ldots,\U_m)\big|\big|\F_j-\G_j\big|\big|_\diamond\\
    = & \sum_{\substack{\U_1,\ldots,\U_m\\j=1,\ldots,m}}p(\U_1,\ldots,\U_m)\big|\big|\E_{c\Z_j,j}-\E_{c\Z_j\U_j,j}\big|\big|_\diamond\:,
\end{align}
where we used the fact that $||\F_j||_\diamond,||\G_j||_\diamond\leq1$ for all $\F_j,\G_j$.
\end{proof}

\begin{figure}[t]
\centering
\includegraphics[scale=0.174]{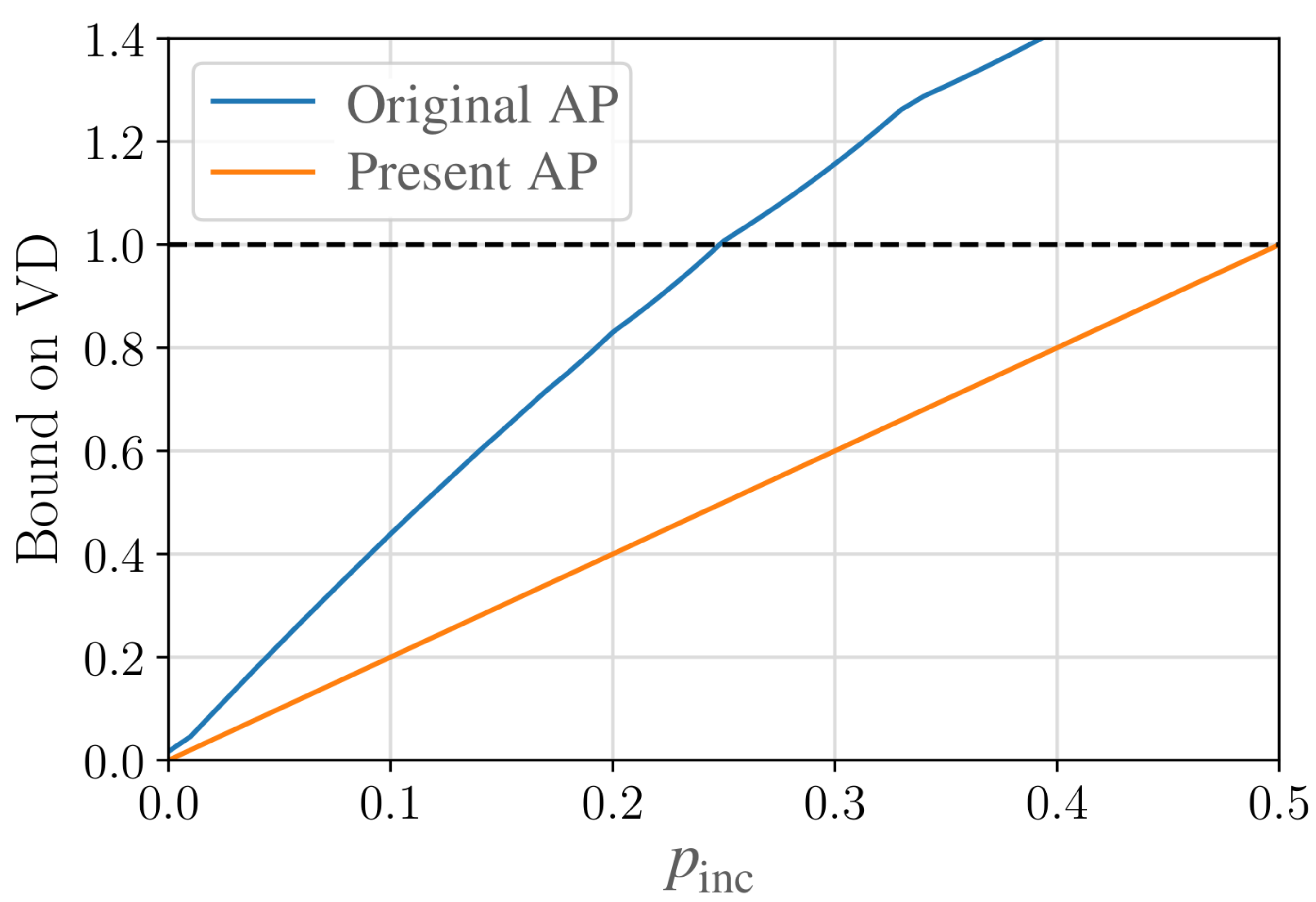}
\caption{\small  The best value of provided by the original AP (equation \ref{eq:best_bound_original}) and the bound provided by the refined AP (r.h.s. of Eq.  \ref{eq:boundVD}) as functions of $p_{\textup{inc}}$.}
\label{fig:comparison_bounds}
\end{figure}

\hypertarget{sec:comparison}{\textit{3. Comparing the present AP with the AP in Ref.~\textup{\cite{FKD18}}|}}In this section we compare the AP demonstrated in this paper (which we name ``present AP'') with the AP in Ref. \cite{FKD18} (which we name ``original AP''), demonstrating that the present AP leads to a significantly tighter bound on the VD. 

In the original AP the user implements the target circuit together with a $v$ trap circuits, initialized in the same way as the trap circuits in the present AP. After implementing all the circuits, the output of the target circuit is accepted only if all the trap circuits return the correct output, otherwise it is discarded. The main result proven in Ref. \cite{FKD18} is that the VD between the probability distribution of the accepted outputs $\{p^{\textup{acc}}_{\textup{exp}}(\overline{s})\}$ and the ideal probability distribution $\{p_{\textup{ideal}}(\overline{s})\}$ can be bounded as
\begin{equation}
\label{eq:boundVDold}
\frac{1}{2}\sum_{\overline{s}}\big|p_{\textup{ideal}}(\overline{s})-p^{\textup{acc}}_{\textup{exp}}(\overline{s})\big|\leq\frac{\kappa}{(v+1)\textup{prob(acc)}}\textrm{ ,}
\end{equation}
where $\kappa\approx{1.7}$ is a constant and prob(acc) is the probability that the output of the target circuit is accepted (which can be measured by running the AP multiple times with the same target and the same number of traps).

To prove that the present AP leads to a better bound than the original AP, we now rewrite the r.h.s. of Eq.~\ref{eq:boundVDold} as a function of the total probability $p_{\textup{inc}}$ that a trap circuit returns an incorrect output. The probability that all the traps return the correct output is $
\textup{prob(acc)}=(1-p_{\textup{inc}})^v$, which gives
\begin{equation}
\frac{1}{2}\sum_{\overline{s}}\big|p_{\textup{ideal}}(\overline{s})-p^{\textup{acc}}_{\textup{exp}}(\overline{s})\big|\leq\frac{\kappa}{(v+1)(1-p_{\textup{inc}})^v}
\end{equation}
The r.h.s. of the above inequality depends on the number $v$ of traps. All the values obtained at different $v$ are valid upper bounds on the VD. The smallest value
\begin{equation}
\label{eq:best_bound_original}
\eta_\textrm{best}=\min_v\frac{\kappa}{(v+1)(1-p_{\textup{inc}})^v}
\end{equation}
corresponds to the best upper bound and can be calculated by implementing the AP many times for different values of $v$.

In Fig. \ref{fig:comparison_bounds} we plot the bounds provided by present AP and original AP as functions of $p_{\textup{inc}}$. As it can be seen, for all the values of $p_{\textup{inc}}$ the latter bound is larger than the former one approximately by a factor 2. Moreover, the bound provided by the original AP exceeds unity for all $p_{\textup{inc}}\gtrsim0.25$, while that provided by the present AP only exceeds unity for $p_{\textup{inc}}\geq0.5$.

While the present AP yields tighter bounds on the VD, the original AP has been proven to be robust to a more general noise model. Indeed, the noise model assumed in Ref. \cite{FKD18} encompasses arbitrary coupling between system and environment, allowing for time-correlated noise. There is thus a trade-off between the generality of noise models captured and the tightness of the bounds obtained.

\end{document}